\documentclass[final,1p,times]{elsarticle}

\usepackage{color}
\usepackage{amsmath}
\usepackage{subfigure}
\usepackage{hyperref}
\usepackage{orcidlink}

\usepackage{tcolorbox}
\usepackage{colortbl}  % 确保已加载此包
\usepackage{gensymb}

\journal{Physics of the Dark Universe}
\makeatletter
\newcommand{\Hunit}{km~s^{-1}~Mpc^{-1}}

\providecommand*{\jnl@style}{\textrm}
\providecommand*{\ref@jnl}[1]{{\jnl@style{#1}}}

\providecommand{\pra}{\ref@jnl{Phys. Rev. A}}
\providecommand{\prb}{\ref@jnl{Phys. Rev. B}}
\providecommand{\prc}{\ref@jnl{Phys. Rev. C}}
\providecommand{\prd}{\ref@jnl{Phys. Rev. D}}
\providecommand{\pre}{\ref@jnl{Phys. Rev. E}}
\providecommand{\prl}{\ref@jnl{Phys. Rev. Lett.}}

\providecommand{\physrep}{\ref@jnl{Phys. Rep.}}
\providecommand{\physscr}{\ref@jnl{Phys. Scr.}}
\providecommand{\jcp}{\ref@jnl{J. Chem. Phys.}}
\providecommand{\physa}{\ref@jnl{Nucl. Phys. A}}
\providecommand{\jcap}{\ref@jnl{JCAP}}
\providecommand{\ap}{\ref@jnl{Astropart. Phys.}}

\providecommand{\apj}{\ref@jnl{ApJ}}
\providecommand{\apjl}{\ref@jnl{ApJL}}
\providecommand{\apjs}{\ref@jnl{ApJS}}
\providecommand{\mnras}{\ref@jnl{MNRAS}}
\providecommand{\aap}{\ref@jnl{A\&A}}
\providecommand{\lrr}{\ref@jnl{Living Rev. Relativ.}}
\providecommand{\grg}{\ref@jnl{Gen. Relativ. Gravit.}}

\providecommand{\aj}{\ref@jnl{AJ}}
\providecommand{\araa}{\ref@jnl{ARA\&A}}
\providecommand{\pasp}{\ref@jnl{PASP}}
\providecommand{\nat}{\ref@jnl{Nature}}
\providecommand{\science}{\ref@jnl{Science}}
\providecommand{\planss}{\ref@jnl{Planet. Space Sci.}}
\makeatother

\definecolor{revisionblue}{RGB}{0,0,255}  % Blue for extensive changes/additions
\definecolor{revisiongreen}{RGB}{0,128,0} % Green for moderate changes
\definecolor{revisionred}{RGB}{255,0,0}   % Red for deletions or corrections
\definecolor{revisiondark}{RGB}{0,0,0}   % Red for deletions or corrections

\newif\ifrevision
\revisiontrue  % Set to \revisionfalse for final version without marks

\begin{document}
	
	\begin{frontmatter}
		
		\title{FLRW Kinematic-Induced Measurement of the Hubble Constant from Cosmic Chronometer and Redshift Drift Observations}

		\author[address1]{Kang Jiao~\orcidlink{0000-0003-0167-9345}}

		\author[address2,address3]{Tong-Jie Zhang~\orcidlink{0000-0002-3363-9965}~\corref{correspondingauthor}}
		\cortext[correspondingauthor]{Corresponding authors}
		\ead{tjzhang@bnu.edu.cn}
		
		\author[address1,address2]{Liang Gao~\orcidlink{0000-0002-9276-917X}~\corref{correspondingauthor}}
		\ead{lgaozzu@zzu.edu.cn}
		
		\author[address4,address5]{Yun Chen~\orcidlink{0000-0001-8919-7409}}
		
		\address[address1]{Institute for Astrophysics, School of Physics, Zhengzhou University, Zhengzhou 450001, China}
		\address[address2]{Institute for Frontiers in Astronomy and Astrophysics, Beijing Normal University, Beijing 102206, China}
		\address[address3]{School of Physics and Astronomy, Beijing Normal University, Beijing 100875, China}
		\address[address4]{National Astronomical Observatories, Chinese Academy of Sciences, Beijing 100101, China}
		\address[address5]{College of Astronomy and Space Sciences, University of Chinese Academy of Sciences, Beijing, 100049, China}
		
		\begin{abstract}

			We present a geometric embedding method that exploits the exact kinematic relation $\dot{z} = H_0(1 + z) - H(z)$ to transform redshift misalignment between Cosmic Chronometer (CC) and Sandage-Loeb (SL) datasets into fundamental constraints in observable space. The approach recognizes that $H_0$ encodes the orientation of the FLRW observational plane defined by $(z, H(z), \dot{z})$ coordinates, enabling direct algebraic determination without parametric assumptions or interpolation schemes.
			Validation using available CC measurements and forecasted redshift drift data from FAST, CHIME, SKA, and ELT demonstrates 1.9\% precision for optimal data combinations, yielding $H_0 = 66.26 \pm 1.26$ km s$^{-1}$ Mpc$^{-1}$ while maintaining complete cosmological model independence. While no actual SL measurements currently exist, requiring us to rely on simulations for validation, our geometric constraints show superior resilience against sparse redshift coverage compared to Gaussian Process (GP) methods, which exhibit systematic biases and large uncertainties when datasets lack substantial overlap.
			This kinematic framework establishes geometric embedding as a robust tool for precision cosmological measurements, offering a fundamentally different approach to $H_0$ determination through pure observational analysis based on FLRW kinematic principles. The full potential of this method awaits implementation with real SL measurements from next-generation facilities.

		\end{abstract}
		
		
	\end{frontmatter}
	
	% \linenumbers

\section{Introduction} 

The Hubble constant $H_0$ tension—now exceeding $5\sigma$—represents one of the most significant challenges to our cosmological framework. This discrepancy manifests as a clear division between early and late-Universe measurements: observations of the early Universe through the Cosmic Microwave Background (CMB) and Baryon Acoustic Oscillation (BAO) yield $H_0 = 67.4 \pm 0.5 \,\mathrm{km\,s^{-1}\,Mpc^{-1}}$~\citep{2020A&A...641A...6P}, while late-Universe measurements using Cepheid-calibrated supernovae consistently give higher values around $H_0 = 73.04 \pm 1.04 \,\mathrm{km\,s^{-1}\,Mpc^{-1}}$~\citep{2022ApJ...934L...7R}. This stark division extends across multiple measurement techniques, with strong-lensing time delays~\citep{2020MNRAS.498.1420W} and megamasers~\citep{2020ApJ...891L...1P} generally aligning with the higher values, while tip of the red giant branch calibrations yield intermediate results of $69$--$70 \,\mathrm{km\,s^{-1}\,Mpc^{-1}}$~\citep{2019ApJ...882...34F}. Recent James Webb Space Telescope (JWST) observations of Cepheids and Type Ia supernovae suggest a potential $\sim 1$--$2 \,\mathrm{km\,s^{-1}\,Mpc^{-1}}$ downward revision in these late-Universe measurements~\citep{2024arXiv240806153F}, though this shift remains insufficient to fully resolve the tension. The persistence of this discrepancy suggests either undiscovered systematic effects in our measurement techniques or, more intriguingly, new physics beyond the standard $\Lambda$CDM model~\citep{2021CQGra..38o3001D}.

In this context, model-independent measurement approaches become crucial for resolving the Hubble tension. Cosmic Chronometers (CC), which estimate the expansion rate \( H(z) \) from the differential aging of passively evolving galaxies, generally favor lower \( H_0 \) values aligned with CMB constraints~\citep{2022LRR....25....6M}. Meanwhile, the Sandage-Loeb (SL) effect offers a unique, model-independent means to directly detect cosmic acceleration by measuring the secular drift in the redshifts of distant sources~\citep{Sandage1962,1998ApJ...499L.111L}. These two approaches are fundamentally linked through the cosmic expansion kinematic relation \citep{Weinberg:2008zzc} derived from the FLRW metric:
\begin{equation}
	- (1+z)H_0 + H(z) + \dot{z}  = 0,                       \label{eq:FLRW}
\end{equation}
where $\dot{z}$ represents the time derivative of redshift (redshift drift). This equation encapsulates how the observable quantities $H(z)$ and $\dot{z}$ directly constrain $H_0$ without requiring assumptions about dark energy or other cosmological parameters.

The SL signal can be probed with increasing promise using the Ly\(\alpha\) forest in quasar spectra at high redshift (\( z \sim 2\text{--}5 \)), especially with next-generation facilities like the ArmazoNes high Dispersion Echelle Spectrograph (ANDES) spectrograph on the Extremely Large Telescope (ELT) that will enable unprecedented precision through its high-stability wavelength calibration system \citep{2008MNRAS.386.1192L,2024ExA....57....5M}, as well as 21-cm radio absorption lines at lower redshift (\( z \sim 0.1\text{--}2.5 \))~\citep{2012ApJ...761L..26D, 2014PhRvL.113d1303Y, 2015aska.confE..27K, 2020JCAP...01..054J, 2020MNRAS.492.2044C}. The combination of these advancing SL measurements with CC data therefore presents a powerful, fully empirical route toward reconstructing the cosmic expansion history and providing an independent and robust determination of \( H_0 \).

However, applying the kinematic relation (Eq.~\ref{eq:FLRW}) in a model-independent way presents significant methodological challenges due to the mismatched redshift distributions between CC and SL measurements. Existing approaches to implement this kinematic constraint face critical limitations:
 (1) Cosmographic techniques using e.g. Taylor or Padé expansions avoid dark energy assumptions but introduce truncation artifacts and convergence issues at higher redshifts \citep{2019GReGr..51....2C}; (2) Nonparametric methods like Gaussian Process (GP) regression offer flexibility but remain sensitive to kernel choice, which implicitly imposes smoothness priors that may not reflect true cosmic evolution \citep{2012PhRvD..85l3530S}; (3) Neural networks can model complex relationships but function as black boxes that limit physical interpretability of results \citep{2024ApJS..270...23Z}; and (4) Binning techniques provide simple interpolation but introduce dependencies on bin width and placement that can systematically affect results and obscure true variations in sparsely sampled redshift regimes \citep{2017APh....86...41L, 2017MNRAS.471L..82M}. These interpolation and smoothing approaches all rely on assumptions about data behavior between measurement points that potentially introduce biases, particularly in the crucial redshift ranges where measurements are sparse.

In this article, we present a novel geometric embedding methodology for model-independent determination of the Hubble constant $H_0$. Unlike existing kinematic estimators that typically rely on parametric reconstruction at redshift $z=0$, our approach directly utilizes the intrinsic kinematic relation between $H(z)$ and $\dot{z}$. While other model-independent methods are often limited to individual probes with their inherent systematics, our framework naturally synergizes multiple measurements without introducing the additional assumptions typically required to address redshift misalignment. This geometric formulation preserves the statistical integrity of each dataset while allowing their mutual constraints to effectively mitigate systematic biases.

In Section~\ref{sec:methodology}, we introduce the geometric embedding methodology that transforms redshift misalignment between CC and SL datasets into fundamental constraints in observable space. Section~\ref{sec:feasibility} demonstrates our approach's feasibility and precision under realistic observational conditions. Sections~\ref{sec:cc_data} to ~\ref{sec:noise} describe the current CC compilation and forecasted SL measurements used for validation. Section~\ref{sec:validation} presents comprehensive validation through comparison with conventional $\Lambda$CDM fitting and GP reconstruction. Section~\ref{sec:discussions} discusses the advantages of kinematic constraints and implications for precision cosmology. Finally, Section~\ref{sec:conclusion} presents our conclusions regarding the effectiveness of geometric methods for high-precision $H_0$ determination.

\section{Methodology}
\label{sec:methodology}
We introduce a geometric approach that transforms the redshift misalignment problem into a fundamental constraint in observable space. The Friedmann--Lema\^{i}tre--Robertson--Walker (FLRW) kinematic relation described in Eq.~\ref{eq:FLRW} implies that all valid cosmic expansion histories must satisfy a specific geometric condition—effectively defining a plane in \((z, H(z), \dot{z})\) space whose orientation directly encodes \(H_0\). This insight allows us to combine CC and SL measurements without imposing model-dependent interpolation schemes or arbitrary priors, offering a robust methodology that relies only on basic kinematic principles. (The visualization of this geometric embedding with real data is shown later in Figure~\ref{fig:geometric_embedding}.) Our inference involves fitting this geometric plane to the combined observational data.

To quantify the deviation of each measurement from the best-fit plane, we define the residual \(r_i\) for each redshift \(z_i\) as follows:
\begin{align}
	r_i &= \tilde{S}_i - (1 + z_i) \bar{H_0},
	\label{eq:residual}
\end{align}
where \(\tilde{S}_i\) denotes the kinematically-constrained hybrid summation that depends on which measurement is available at redshift $z_i$:
\begin{align}
	\tilde{S}_i = 
	\begin{cases}
		H(z_i) + \tilde{\dot{z}}_i, & \text{when $H(z_i)$ is measured} \\
		\tilde{H}(z_i) + \dot{z}_i, & \text{when $\dot{z}_i$ is measured}
	\end{cases}.
	\label{eq:embed}
\end{align}
The tilde symbol denotes quantities treated as free variables to be embedded, while non-tilde quantities represent directly observed values. While these embedded values may appear to introduce flexibility, the only true degree of freedom in the system remains $H_0$. Without imposing global consistency across redshifts, $H_0$ would indeed be arbitrary when considering each redshift point independently.

Additionally, $\bar{H_0}$ is the inverse variance-weighted mean of the $H_0$ estimates at each redshift, $\tilde{H}_{0,i} = \tilde{S}_i / (1 + z_i)$, with weights from error propagation. The weighted mean is computed as:
\begin{equation}
\bar{H_0} = \frac{\sum_i w_i \tilde{H}_{0,i}}{\sum_i w_i},
\end{equation}
where the weights are $w_i = 1/\sigma_{\tilde{H}_{0,i}}^2$, with $\sigma_{\tilde{H}_{0,i}}$ representing the propagated uncertainty in each individual $H_0$ estimate. The error propagation from $\tilde{S}_i$ to $\tilde{H}_{0,i}$ follows:
\begin{equation}
\sigma_{\tilde{H}_{0,i}}^2 = \left(\frac{\partial \tilde{H}_{0,i}}{\partial \tilde{S}_i}\right)^2 \sigma_{\tilde{S}_i}^2 = \frac{\sigma_{\tilde{S}_i}^2}{(1 + z_i)^2},
\end{equation}
where $\sigma_{\tilde{S}_i}^2$ represents the combined variance of $\tilde{S}_i$. Since observed values and embedded parameters are statistically independent, this variance follows:
\begin{align}
	\sigma_{\tilde{S}_i}^2 = 
	\begin{cases}
		\sigma_{H(z_i)}^2 + \sigma_{\tilde{\dot{z}}_i}^2, & \text{when $H(z_i)$ is measured} \\
		\sigma_{\tilde{H}(z_i)}^2 + \sigma_{\dot{z}_i}^2, & \text{when $\dot{z}_i$ is measured}
	\end{cases}.
\end{align}
These expressions encompass both observational uncertainties and systematic effects. The inverse variance weighting optimally combines measurements with heterogeneous precisions, automatically emphasizing high-precision observations while preserving information from less precise data points.

Each redshift yields an independent $H_0$ estimate, collectively defining a statistical plane in the observational space. The weighted mean $\bar{H_0}$ has variance $\text{Var}(\bar{H_0}) = 1/\sum_i w_i$, achieving the minimum possible uncertainty for any linear combination of the individual estimates. Alignment with this plane results in residuals $r_i = \tilde{H}_{0,i} - \bar{H_0}$ with expectation value $\langle r_i \rangle = 0$, indicating robustness against systematic biases that might affect individual measurements differently. The inverse variance weighting enhances this reliability by prioritizing lower-uncertainty observations while maintaining statistical optimality. Significant non-zero residuals suggest discrepancies between observed and predicted values, necessitating further examination of systematic errors or assumptions.

The covariance matrix associated with the residual vector $\boldsymbol{r}$ propagates from observational uncertainties and $\bar{H_0}$ scatter:
\[
\boldsymbol{C}=  
\begin{bmatrix}  
\boldsymbol{C}_{\mathrm{CC}} & 0 \\[4pt]  
0 & \boldsymbol{C}_{\mathrm{SL}}  
\end{bmatrix} + \text{Var}(\bar{H_0}) \cdot (1+\boldsymbol{z})(1+\boldsymbol{z})^{\mathrm{T}},
\]
where $\boldsymbol{C}_{\mathrm{CC}}$ and $\boldsymbol{C}_{\mathrm{SL}}$ represent the respective covariance matrices for CC and SL observations. Explicitly, the CC covariance matrix follows~ \citep{2020ApJ...898...82M}:
\[
(\boldsymbol{C}_{\mathrm{CC}})_{ij} = 
(\mathrm{Cov}_{ij}^\mathrm{stat} 
+ \mathrm{Cov}_{ij}^\mathrm{young} 
+ \mathrm{Cov}_{ij}^\mathrm{model} 
+ \mathrm{Cov}_{ij}^\mathrm{met}),
\]
where terms represent contributions from statistical errors, young stellar component contamination, model dependencies, and metallicity effects. The model-dependent component further decomposes as:
\[
\mathrm{Cov}_{ij}^\mathrm{model}
= \mathrm{Cov}_{ij}^{\mathrm{SFH}} + \mathrm{Cov}_{ij}^{\mathrm{IMF}} + \mathrm{Cov}_{ij}^{\mathrm{st.lib.}} + \mathrm{Cov}_{ij}^{\mathrm{SPS}}.
\]
For SL measurements, we assume $(\boldsymbol{C}_{\mathrm{SL}})_{ij} = \delta_{ij} \, \sigma_{\mathrm{SL},i}^2$, indicating uncorrelated uncertainties.  The block-diagonal structure reflects the assumption that CC and SL measurements are statistically independent, as they probe complementary physical processes through distinct observational techniques---stellar population synthesis versus spectroscopic redshift drift detection. Detailed covariance specifications are provided in the section~\ref{sec:cc_data} and ~\ref{sec:forecastedSLdata}.

The tilde symbol quantities ($\tilde{H}$ and $\tilde{\dot{z}}$) represent embedded values that are treated as free parameters in our Bayesian inference framework. 
We employ Markov Chain Monte Carlo (MCMC) methods for Bayesian inference, sampling the posterior distribution $P(\tilde{H}, \tilde{\dot{z}}|\mathrm{data}) \propto \mathcal{L}(\mathrm{data}|\tilde{H}, \tilde{\dot{z}}) \times P(\tilde{H}, \tilde{\dot{z}})$ to achieve robust parameter constraints while naturally accounting for uncertainties in both the measurements and embedded values. With $\boldsymbol{r}$ as the residual vector and $\boldsymbol{C}$ as the covariance matrix, the Gaussian log-likelihood is expressed as  
\begin{equation}  
\ln \mathcal{L} = -\frac{n}{2} \ln(2\pi) - \frac{1}{2} \ln \det(\boldsymbol{C}) - \frac{1}{2} \boldsymbol{r}^{\mathrm{T}} \boldsymbol{C}^{-1} \boldsymbol{r},
\end{equation}
where the terms represent normalization, uncertainty weighting, and model-data agreement, respectively.
In practical terms, the computation proceeds as follows during each MCMC iteration: (1) A proposed set of values for all $\tilde{H}$ and $\tilde{\dot{z}}$ parameters is generated; (2) For each redshift $z_i$, we compute $\tilde{S}_i$ using Eq.~\ref{eq:embed}, combining observed values with the proposed embedded values; (3) Each $\tilde{S}_i$ yields an independent $H_0$ estimate via $\tilde{H}_{0,i} = \tilde{S}_i / (1 + z_i)$; (4) These $\tilde{H}_{0,i}$ values are combined into the weighted mean $\bar{H}_0$ using weights derived only from observational uncertainties for that MCMC step; (5) Residuals $r_i$ are calculated using Eq.~\ref{eq:residual}; and (6) The likelihood is evaluated via the above equation. 
During each MCMC step, all proposed parameter values are treated as fixed points without intrinsic uncertainty, as this is fundamental to correctly sampling the posterior distribution. Only after completing the full MCMC sampling do we incorporate the statistical uncertainties in our final results, where the uncertainties of embedded values ($\sigma_{\tilde{\dot{z}}_i}$ and $\sigma_{\tilde{H}(z_i)}$) are derived from the width of their respective posterior distributions. This distinction ensures that our MCMC procedure correctly explores the parameter space while our final error analysis properly accounts for the full statistical properties of the parameters. This approach enables us to enforce the geometric constraint across all redshifts while obtaining robust uncertainty estimates.

For the implementation, we employ the affine-invariant ensemble sampler \texttt{emcee} \cite{2013ascl.soft03002F} with walkers numbering at least 4 times the parameter dimension (typically 100--152 walkers in our analysis). Our implementation monitors convergence using integrated autocorrelation time estimates, discarding the first 20\% of each chain as burn-in. We utilize a stretch move with scale parameter $a=1.5$ as the primary proposal mechanism, and verify convergence by ensuring the chain lengths significantly exceed the maximum autocorrelation time across all parameters.

Observationally mutually exclusive \(H(z)\) and \(\dot{z}\) measurements create an underdetermined system at each single redshift as outlined in Eq.~(\ref{eq:embed}), which can yield arbitrary $\tilde{H}_{0,i}$ values that satisfy a zero residual condition. These informationally complementary measurements collectively imply a common underlying $\bar{H}_0$ value. We assume plausible boundaries from observations at adjacent redshifts for the same physical quantity to embed, assuming the missing complementary quantity lies within a 5$\sigma$ range obtained through linear interpolation/extrapolation to infer reasonable $H_0$ values. Specifically, for a missing quantity $X$ (either $H(z)$ or $\dot{z}$ at redshifts where only the other one is measured) at redshift $z$,  we estimate its mean $\mu_X(z)$ by interpolation or extrapolation between the nearest available measurements, while rigorously propagating uncertainties and covariances to compute its standard deviation:
\begin{align}
\mu_X(z) &= \mu_X(z_1) + \frac{z-z_1}{z_2-z_1} \left[\mu_X(z_2) - \mu_X(z_1)\right], \\
\sigma_X^2(z) &= \sigma_1^2 + \left[\frac{\sigma_1^2+\sigma_2^2-2\,\mathrm{Cov}_{12}}{(z_2-z_1)^2}\right](z-z_1)^2 + 2\left[\frac{\sigma_1^2-\mathrm{Cov}_{12}}{z_2-z_1}\right](z-z_1).
\end{align}
Here, $\sigma_1^2$ and $\sigma_2^2$ are the variances of $X$ at redshifts $z_1$ and $z_2$ respectively, and $\mathrm{Cov}_{12}$ represents the covariance between the measurements $X(z_1)$ and $X(z_2)$.
For interpolation, $z_1$ and $z_2$ are the nearest redshifts with measured $X$ values such that $z_1 < z < z_2$. For extrapolation at the lower redshift boundary, $z_1$ and $z_2$ are the two lowest redshifts with measured $X$ values (where $z < z_1 < z_2$). Similarly, for extrapolation at the upper boundary, they represent the two highest redshifts with measured $X$ values (where $z_1 < z_2 < z$). A non-informative uniform prior $X(z)\sim \mathcal{U}(\mu_X(z)-5\sigma_X(z),\ \mu_X(z)+5\sigma_X(z))$ is then imposed. The boundaries, shown as gray regions in Figure~\ref{fig:geometric_embedding_results}, encompass all physically plausible values while imposing minimal assumptions, enhancing $H_0$ estimation reliability without introducing strong assumptions about the smoothness.

\section{Feasibility}
\label{sec:feasibility}
	To assess our geometric method's performance under realistic observational conditions, we combine existing CC observations with forecasted SL data. This mixed real-simulated dataset evaluates the precision achievable when applying our approach to current CC constraints and anticipated redshift drift measurements.

% \FloatBarrier
\subsection{Current Available CC Data}
\label{sec:cc_data}
\begin{table}[!htbp]
\caption{Hubble parameter measurements from CCs. Data spans $z = 0.07-1.965$ using various differential age methods: full-spectrum fitting (F), Lick indices (L), $D4000$ index (D), and machine learning (ML).}
\label{tab:cc_measurements}
\centering
% \tiny 
\setlength{\tabcolsep}{8pt}  
\begin{tabular}{lcccccl}
	\\
\hline\hline
$z$ & $H(z)$ & $\sigma_{\mathrm{stat}}$ & $\sigma_{\mathrm{sys}}$ & $\sigma_{\mathrm{to t}}$ & Method & Reference \\
% & (km/s/Mpc) & (km/s/Mpc) & (km/s/Mpc) & (km/s/Mpc) &  & \\
% & \multicolumn{4}{c}{ km s$^{-1}$ Mpc$^{-1}$} & &\\
\hline
0.07 & 69.0 &  &  & 19.6 & F & \cite{2014RAA....14.1221Z} \\
% 0.090 & 69 & 12 & F & \cite{2003ApJ...593..622J} \\
0.09 & \underline{72} &  &  & \underline{13} & F & \cite{2003ApJ...593..622J}$^{\ddagger}$ \\
0.12 & 68.6 &  &  & 26.2 & F & \cite{2014RAA....14.1221Z} \\
0.17 & 83 &  &  & 8 & F & \cite{2005PhRvD..71l3001S}$^{*}$ \\
0.1791 & 75 & (3.8) &  & 4 & D & \cite{2012JCAP...08..006M} \\
0.1993 & 75 & (4.9) &  & 5 & D & \cite{2012JCAP...08..006M} \\
0.20 & 72.9 &  &  & 29.6 & F & \cite{2014RAA....14.1221Z} \\
0.27 & 77 &  &  & 14 & F & \cite{2005PhRvD..71l3001S}$^{*}$ \\
0.28 & 88.8 &  &  & 36.6 & F & \cite{2014RAA....14.1221Z} \\
0.3519 & 83 & (13) &  & 14 & D & \cite{2012JCAP...08..006M} \\
0.3802 & 83.0 & 4.3 & 12.9 & 13.5 & D & \cite{2016JCAP...05..014M} \\
0.4 & 95 &  &  & 17 & F & \cite{2005PhRvD..71l3001S}$^{*}$ \\
0.4004 & 77.0 & 2.1 & 10 & 10.2 & D & \cite{2016JCAP...05..014M} \\
0.4247 & 87.1 & 2.4 & 11 & 11.2 & D & \cite{2016JCAP...05..014M} \\
0.4497 & 92.8 & 4.5 & 12.1 & 12.9 & D & \cite{2016JCAP...05..014M} \\
0.47 & 89 &  23 & 44 & <49.6> & F & \cite{2017MNRAS.467.3239R} \\
0.4783 & 80.9 & 2.1 & 8.8 & 9 & D & \cite{2016JCAP...05..014M} \\
0.48 & 97 &  &  & 62 & F & \cite{2010JCAP...02..008S}$^{**}$ \\
0.5 & 72.1 & 33.9 & 7.3 & <34.7>  & D & \cite{2025MNRAS.540.3135L}\\
0.5929 & 104 & (11.6) &  & 13 & D & \cite{2012JCAP...08..006M} \\
0.6797 & 92 & (6.4) &  & 8 & D & \cite{2012JCAP...08..006M} \\
0.75 & 98.8 & 24.8 &  & 33.6 & L & \cite{2022ApJ...928L...4B}$^{\dagger}$ \\
0.75 & 105.0 & 7.9 & 7.3 & <10.8> & ML & \cite{2023JCAP...11..047J}$^{\dagger}$ \\
0.7812 & 105 & (9.4) &  & 12 & D & \cite{2012JCAP...08..006M} \\
0.8 & 113.1 & 15.1 & $^{+29.1}_{-11.3}$ & <23.6> & F & \cite{2023ApJS..265...48J}$^{\dagger}$ \\
0.8754 & 125 & (15.3) &  & 17 & D & \cite{2012JCAP...08..006M} \\
0.88 & 90 &  &  & 40 & F & \cite{2010JCAP...02..008S}$^{**}$ \\
0.9 & 117 &  &  & 23 & F & \cite{2005PhRvD..71l3001S}$^{*}$ \\
1.037 & 154 & (13.6) &  & 20 & D & \cite{2012JCAP...08..006M} \\
1.26 & 135 & 60 & 27 & 65 & F & \cite{2023AA...679A..96T} \\
1.3 & 168 &  &  & 17 & F & \cite{2005PhRvD..71l3001S}$^{*}$ \\
1.363 & 160 & (23.07) &  & 33.6 & D & \cite{2015MNRAS.450L..16M} \\
1.43 & 177 &  &  & 18 & F & \cite{2005PhRvD..71l3001S}$^{*}$ \\
1.53 & 140 &  &  & 14 & F & \cite{2005PhRvD..71l3001S}$^{*}$ \\
1.75 & 202 &  &  & 40 & F & \cite{2005PhRvD..71l3001S}$^{*}$ \\
1.965 & 186.5 & (35.05) &  & 50.4 & D & \cite{2015MNRAS.450L..16M} \\
\hline\hline
\end{tabular}

\begin{flushleft}
{\footnotesize \textbf{Notes}:  H(z) and uncertainties are given in km s$^{-1}$ Mpc$^{-1}$. Parentheses and angle brackets indicate uncertainties from \href{https://gitlab.com/mmoresco/CCcovariance/data}{CCcovariance Project} and quadrature combinations, respectively. Underscore values are corrected from commonly reported $H_0 = 69 \pm 12~\mathrm{\Hunit}$ extrapolation to actual $H(z = 0.09)$ value following the original \citet{2003ApJ...593..622J}.
All unmarked data are reported at their original precision as presented in the source publications.\\
$^{\dagger}$ Overlapping samples that should not be used simultaneously.  \\
$^{**}$ Measurements based on galaxy samples that encompass all from ref.~\cite{2005PhRvD..71l3001S}, which contains the data marked with $^{*}$ that should be used with caution according to \citet{2023MNRAS.518..585A}.\\
}
\end{flushleft}
\end{table}

The current available CC dataset comprises 36 measurements spanning $z \sim 0.07 - 1.965$, extending the compilation from \citet{2023arXiv230709501M} with a recent measurement using a sample of 53 brightest cluster galaxies (BCGs) selected by \citet{2025MNRAS.540.3135L} from Southern African Large Telescope (SALT) observations.  These measurements employ various differential age methods including full-spectrum fitting (F), Lick indices (L), $D4000$ index (D), and machine learning (ML) techniques to estimate the Hubble parameter at different redshifts. Statistical and systematic uncertainties for measurements after \citet{2010JCAP...02..008S} can be distinguished, while earlier measurements do not separate these uncertainty components. 
We take covariances within $D4000$ measurements according to the \href{https://gitlab.com/mmoresco/CCcovariance/data}{CCcovariance Project}, incorporating correlations  from stellar population contamination \citep{2018ApJ...868...84M}, model dependencies (star formation history, IMF, stellar libraries), and metallicity effects \citep{2020ApJ...898...82M}. Although potential covariances may exist between other measurements, the complex differences in analysis methods make it difficult to estimate these correlations accurately. We therefore assume independence between different measurement methods and datasets, which represents a reasonable approximation given the heterogeneous nature of the data compilation, though this simplification may underestimate the true uncertainties and could potentially affect the precision of our $H_0$ determination.

We note that the measurement of \citet{2003ApJ...593..622J}, commonly reported as $H(z = 0.09) = 69 \pm 12~\mathrm{\Hunit}$, actually represents an $H_0$ value extrapolated from $z_{\mathrm{eff}} = 0.09$. The correct value at the measurement redshift is $H(z = 0.09) = 72 \pm 13~\mathrm{\Hunit}$, derived by inverting the original $H_0$ determination procedure.
The three datasets marked with $\dagger$ in Table~\ref{tab:cc_measurements} share overlapping samples, and we selected the measurement with the highest precision for our analysis to avoid potential correlations between these overlapping datasets. We also exclude the original measurements from \citet{2005PhRvD..71l3001S}, despite their widespread use in the CC literature. This decision is driven by two key concerns. Subsequent observations by \citet{2010JCAP...02..008S} using more than the same galaxy samples reported fewer data points with significantly larger uncertainties, casting doubt on the reliability of the earlier dataset. Moreover, the comprehensive analysis of \citet{2023MNRAS.518..585A} explicitly advises against utilizing these measurements, further supporting our choice to rely on more robust data. After these careful selection, our final dataset comprises 26 measurements, ensuring that our geometric embedding method is grounded in the most reliable and extensive CC data currently available.

\subsection{Forecasted SL Data}
\label{sec:forecastedSLdata}
For SL simulations, we employ predicted precisions from ongoing and upcoming radio surveys, using the Five-hundred-meter Aperture Spherical Telescope (FAST) \citep{2020JCAP...01..054J}, Canadian Hydrogen Intensity Mapping Experiment (CHIME) \citep{2014PhRvL.113d1303Y}, Square Kilometre Array (SKA) \citep{2015aska.confE..27K} and optical Extremely Large Telescope (ELT) \citep{2008MNRAS.386.1192L} survey following \citet{2022LRR....25....6M}. 

FAST has demonstrated initial capabilities for direct cosmic acceleration measurements through HI 21-cm absorption spectroscopy, achieving 10-100 Hz spectral resolution and signal-to-noise ratios of 57 in 10-minute observations \citep{2022PDU....3701088L, 2024RAA....24g5002K}, with theoretical models predicting $\sim$800 detectable systems via CRAFTS surveys \citep{2020JCAP...01..054J}. However, the current redshift drift precision of the $\sim 10^{-7}$ decade$^{-1}$ remains three orders of magnitude above the cosmologically competitive target of $\sim 10^{-10}$ decade$^{-1}$, representing a substantial technical challenge that requires fundamental advances in systematic error control and long-term stability. Although direct cosmological constraints may remain elusive in the near term, FAST's intermediate-redshift coverage ($z \sim 0-0.35$) and demonstrated spectroscopic capabilities establish it as a valuable testbed for developing the methodologies and technical infrastructure necessary for future high-precision redshift drift measurements with next-generation facilities.

CHIME's unique cylindrical design enables continuous monitoring of the northern sky, offering distinct advantages for redshift drift surveys. The telescope can track thousands of HI absorption systems simultaneously across $z \sim 0.8-2.5$, with the potential to accumulate multi-year baselines through daily observations \citep{2014PhRvL.113d1303Y}. This approach transforms the traditional challenge of coordinating sparse multi-epoch observations into a systematic monitoring program, where statistical uncertainties decrease through extended temporal coverage rather than requiring dramatically increased collecting area.

SKA represents the next generation of radio interferometry with transformative potential for cosmological redshift measurements. Our modeling focuses specifically on absorption line spectroscopy, where SKA's frequency stability and spectral resolution provide optimal conditions for velocity centroiding. We simulate a targeted program monitoring HI absorption systems in two redshift windows: intermediate redshift ($z = 0.55$) and higher redshift ($z = 0.85$) bins. Each bin contains 500 carefully selected absorption systems, chosen for optimal signal characteristics and sky distribution. The program assumes a 10-year monitoring baseline to accumulate sufficient temporal leverage for detecting cosmological acceleration signatures. Under these observational parameters, we model rms uncertainties in cosmic acceleration of approximately 0.08~cm~s$^{-1}$~yr$^{-1}$ according to \citet{2022LRR....25....6M}, reflecting SKA's enhanced sensitivity and spectral capabilities compared to current-generation facilities.

For ELT, we adopt the COsmic Dynamics EXperiment (CODEX) project specifications for measuring redshift drift through Lyman-$\alpha$ forest observations. We simulate 5 data points uniformly distributed in $z = 2-5$, with 30 QSOs per redshift bin. The velocity measurement uncertainty follows the scaling relation from Monte Carlo simulations~\citep{2008MNRAS.386.1192L}:
\begin{equation}
\sigma_v = 1.4 \left(\frac{S/N}{2350}\right)^{-1} \left(\frac{N_{\rm QSO}}{30}\right)^{-1/2} \left(\frac{1+z_{\rm QSO}}{5}\right)^{-\lambda} \, \text{cm s}^{-1},
\end{equation}
where $\lambda = 1.7$ up to $z = 4$ and $\lambda = 0.9$ for $z > 4$. Given $\Delta v/\Delta t_0 = c\dot{z}/(1+z)$, the redshift drift uncertainty becomes:
\begin{equation}
\sigma_{\dot{z}} = \frac{\sigma_v (1+z)}{c \Delta t_0}.
\end{equation}
We assume $S/N = 3000$ per pixel and a 10-year monitoring baseline. While our analysis follows this formulation, we note important recent developments relevant to the forthcoming ANDES ultra-stable spectrograph on the ELT (which evolved from the earlier CODEX concept). Notably, the QUBRICS survey has identified a "Golden Sample" \citep{2023MNRAS.522.2019C} of significantly brighter quasars in the redshift range $2.9 < z < 4.8$, which enables more optimal target selection for redshift drift measurements. Combined with the advanced design of ANDES \citep{2024ExA....57....5M}, these improvements are projected to substantially reduce the total integration time needed for a $3\sigma$ detection of cosmological redshift drift—from approximately 4000 hours (CODEX-era estimates) to less than 2500 hours over a 25-year period~\citep{2020ApJS..250...26B}. This increased efficiency preserves the fundamental measurement approach used in our analysis and demonstrates the enhanced feasibility of future experiments.

	\begin{table}[!htbp]
	\caption{Redshift drift simulations from various radio and optical telescope observations.}
	\label{tab:redshift_drift}
	\centering
	\setlength{\tabcolsep}{8pt}  % 调整列间距
	\begin{tabular}{lcccl}
	\hline\hline
	$z$ & $\sigma_{\Delta v}~\rm(cm/s) $ & $\sigma_{\dot{z}}~\rm(km/s/Mpc)$ & Observations & Reference \\
	\hline
	0.25 & 0.77 & 3.13 & FAST & \citep{2020JCAP...01..054J}\\
	0.55 & 0.29 & 1.49 & SKA Absorption & \citep{2015aska.confE..27K}\\
	0.85 & 0.27 & 1.65 & SKA Absorption & \citep{2015aska.confE..27K}\\
	1.01 & 0.77 & 5.04 & CHIME & \citep{2014PhRvL.113d1303Y}\\
	1.44 & 0.88 & 6.98 & CHIME & \citep{2014PhRvL.113d1303Y}\\
	1.86 & 1.30 & 12.11 & CHIME & \citep{2014PhRvL.113d1303Y}\\
	2.29 & 1.39 & 14.88 & CHIME & \citep{2014PhRvL.113d1303Y}\\
	2.00 & 2.54 & 24.87 & ELT & \citep{2008MNRAS.386.1192L}\\
	2.80 & 1.70 & 21.08 & ELT & \citep{2008MNRAS.386.1192L}\\
	3.50 & 1.28 & 18.72 & ELT & \citep{2008MNRAS.386.1192L}\\
	4.20 & 1.03 & 17.46 & ELT & \citep{2008MNRAS.386.1192L}\\
	5.00 & 0.91 & 17.71 & ELT & \citep{2008MNRAS.386.1192L}\\
	\hline\hline
	\end{tabular}
	\end{table}

Table~\ref{tab:redshift_drift} consolidates these forecasted measurement precisions across the various observational platforms. The combined dataset spans a comprehensive redshift range from $z=0.25$ to $z=5.0$, capturing the predicted transition from positive to negative redshift drift that occurs at intermediate redshifts in standard cosmological models. For our analysis, we adopt the measurement uncertainties as presented in the table, which represent the anticipated technical capabilities of each facility based on their respective observational strategies and instrumental characteristics.

While our formalism accommodates full covariance structures, we adopted a diagonal $\boldsymbol{C}_{\mathrm{SL}}$ for current forecasts, reflecting that each drift measurement derives from independent spectral observations. We adopted a fiducial flat-$\Lambda$CDM cosmology with parameters ($\Omega_m = 0.344$, $H_0 =66.17 $ km s$^{-1}$ Mpc$^{-1}$) derived from a preliminary fit to the CC data itself. All simulated data points are generated based on this fiducial model, avoiding external priors from other measurements. This self-consistent approach ensures an unbiased assessment of the method's intrinsic constraining power on $H_0$ by maintaining internal consistency between the datasets.

\subsection{Noise Simulation}
\label{sec:noise}
Having established our fiducial simulation, we next address the critical issue of noise simulation for method validation. We generate synthetic observational noise by drawing from the multivariate Gaussian distribution defined by our covariance matrix $\boldsymbol{C}_{\mathrm{SL}}$. However, this standard approach introduces an important statistical consideration: each noise realization represents a single random draw from the underlying distribution, leading to inherent variability in parameter recovery that is independent of the analysis methodology itself.

	\begin{figure*}[htbp]
		\centering
		\begin{minipage}[b]{0.32\textwidth}
			\centering
			\hspace{5mm} Noiseless Data\\[0.5em]
			\includegraphics[width=\textwidth]{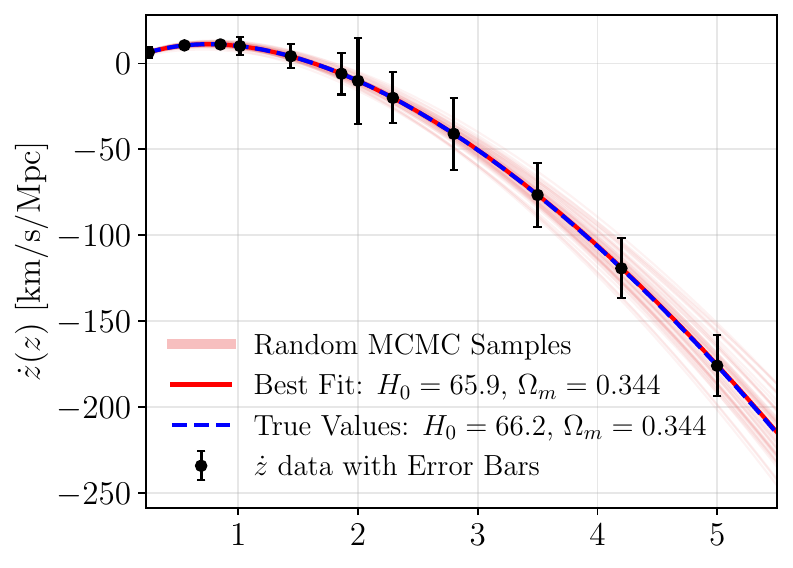}
		\end{minipage}
		\hfill
		\begin{minipage}[b]{0.32\textwidth}
			\centering
			\hspace{5mm} Random Noisy Data\\[0.5em]
			\includegraphics[width=\textwidth]{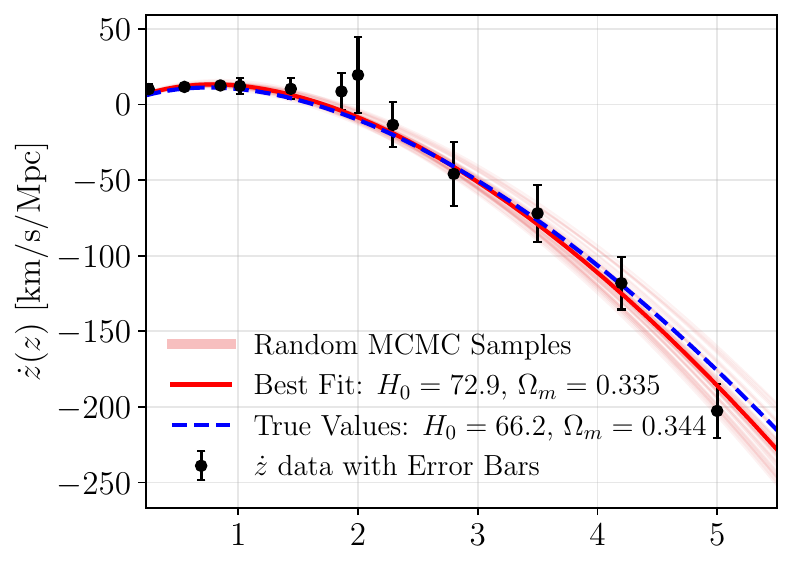}
		\end{minipage}
		\hfill
		\begin{minipage}[b]{0.32\textwidth}
			\centering
			\hspace{5mm} Optimized Noisy Data\\[0.5em]
			\includegraphics[width=\textwidth]{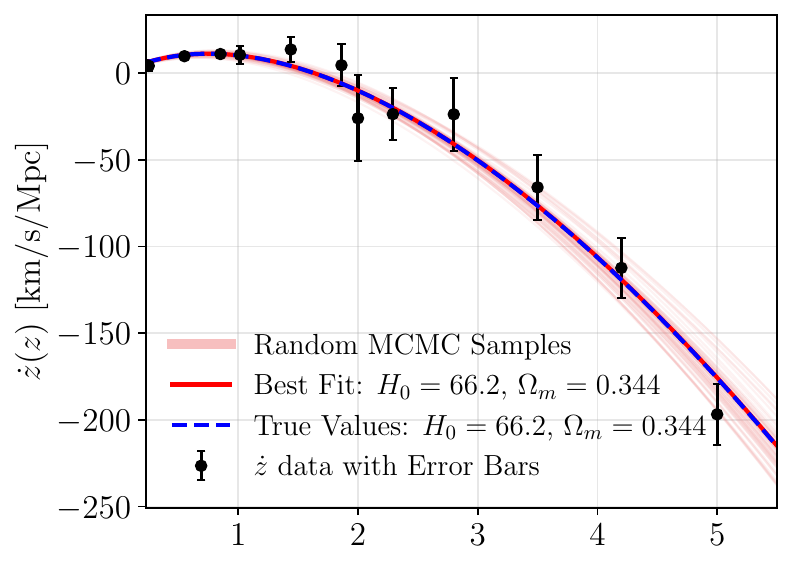}
		\end{minipage}
		
		\caption{Comparison of fiducial model fitting to simulated SL redshift drift data under different noise realizations across redshift range $z = 0 - 5$. The three columns represent noiseless theoretical data (left), random noise realization (middle), and optimized noise seed selection (right). Each panel displays the redshift drift signal $\dot{z}(z)$ with random MCMC samples (shaded regions), best-fit fiducial models (solid lines), true fiducial values (dashed lines), and simulated data points with error bars. This comparison demonstrates the impact of noise seed optimization on fiducial cosmological parameter recovery accuracy.}
		\label{fig:nosie_patterns}
	\end{figure*}

Figure~\ref{fig:nosie_patterns} demonstrates this phenomenon using simulated SL redshift drift data across three noise scenarios. The noiseless case (left panel) establishes our theoretical baseline, where the fitting procedure recovers the input parameters exactly. When realistic measurement uncertainties are introduced, however, individual parameter estimates scatter around the true values. The middle panel shows what can happen with an unlucky noise draw. The recovered Hubble constant ($H_0 = 72.88^{+4.01}_{-3.99}$~km~s$^{-1}$~Mpc$^{-1}$) deviates substantially from the true value ($66.17$~km~s$^{-1}$~Mpc$^{-1}$), not because our methodology fails, but because this particular noise realization pushes the fit away from the correct answer. Such outcomes occur with finite probability even when the analysis is statistically sound.

This reflects a fundamental property of parameter estimation: while individual measurements vary randomly, their average converges to the true values with sufficient sampling. Each estimate $\hat{\theta}_i$ represents a random draw from a distribution centered on $\theta_0$, ensuring that $\mathbb{E}[\hat{\theta}_i] = \theta_0$ when averaged over all possible noise patterns.

For method validation, this creates a practical problem as we need to distinguish genuine systematic errors from random scatter in finite samples. Our solution is to use optimized noise configurations that minimize parameter recovery offsets when applying Bayesian inference with the simulation's underlying model. Such an optimization approach simply identifies realizations near this distribution's center rather than its tails, allowing us to validate the method without compromising statistical integrity. For computational efficiency, we divide our dataset into subgroups and apply different random seeds to each. Our algorithm then searches through seed combinations to find noise realizations where recovered parameters closely match true values. We progressively fix seeds that reduce parameter offsets (measured in $\sigma$ units) below threshold values, typically achieving 50\% reduction from previous baselines. This iterative process ultimately achieves parameter deviations better than $0.5^N\sigma$ where $N=6$ is our number of subgroups. Multiple noise configurations can yield accurate parameter recoveries, consistent with our earlier statement that individual estimates scatter randomly while statistically unbiased. By selecting a configuration with acceptable offset (i.e. offset $< 0.05 \sigma$ as an early stop condition), we validate the method's capacity to recover true parameters under representative noise conditions. The right panel of Figure~\ref{fig:nosie_patterns} shows such a case, where the estimated Hubble constant ($H_0 = 66.19 \pm 4.00$~km~s$^{-1}$~Mpc$^{-1}$) closely matches the input value, in contrast to the substantial deviation with a random noise realization in the middle panel.

Importantly, this simulation does not constitute cherry-picking of favorable results. Since our parameter recovery is statistically unbiased, demonstrating these central realizations provide appropriate test data with noise properties that enable effective method validation. When applying our embedding method to real observational data, we would of course work with the actual noise present in the measurements, with the confidence only if our proposed new methodology has been validated to produce unbiased parameter estimates. The comparison across noise realizations also quantifies how sensitive cosmological constraints are to measurement uncertainties--information that proves valuable when interpreting tensions between different experiments.

\subsection{Comparison and Validation}
\label{sec:validation}

\begin{figure*}[htbp!]
			\centering
			\begin{minipage}[b]{0.45\textwidth}
				\centering
				\hspace{5mm} CC Data\\[0.5em]
				\includegraphics[width=0.9\textwidth]{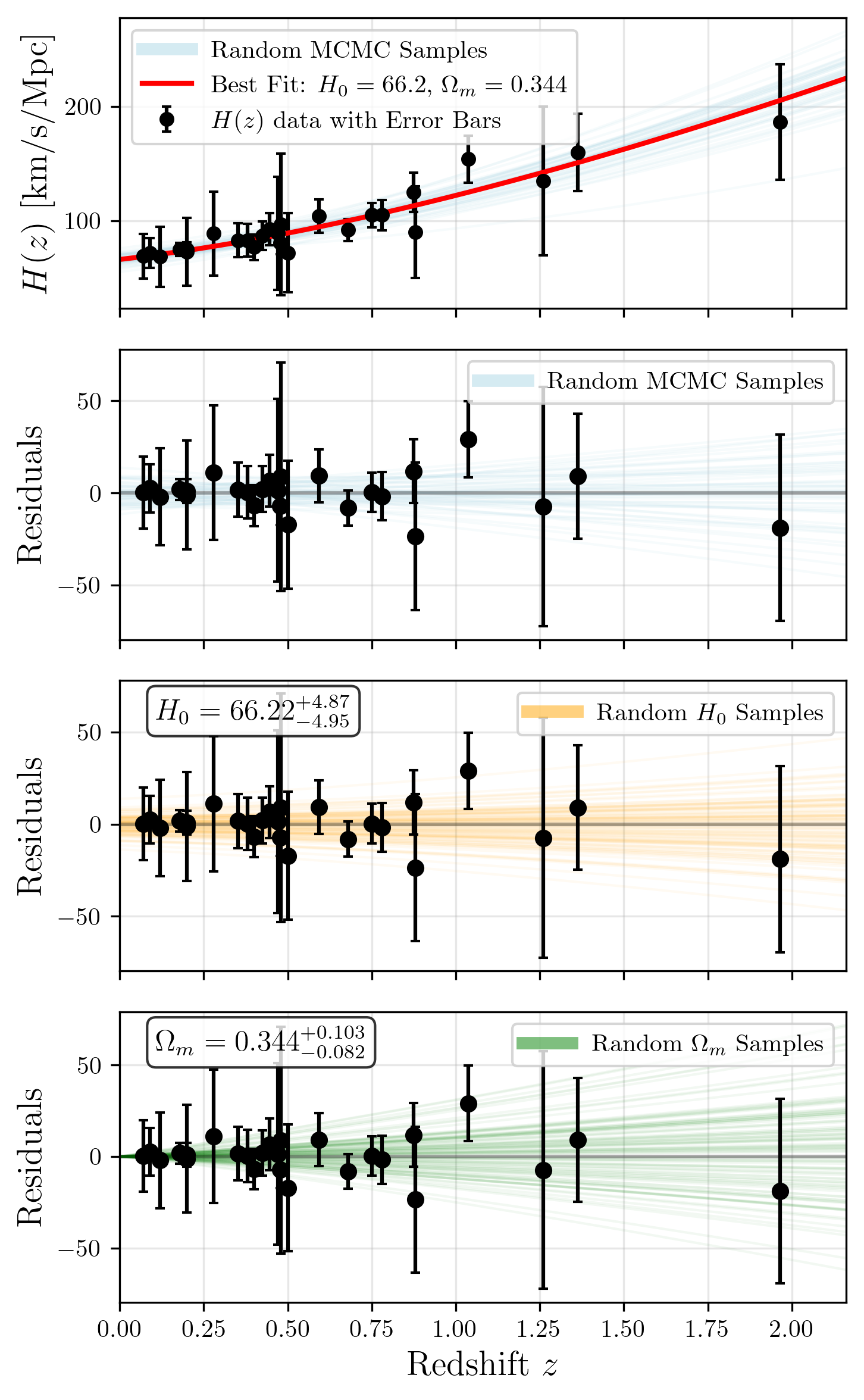} % Reduced width
			\end{minipage}
			\hfill
			\begin{minipage}[b]{0.45\textwidth}
				\centering
				\hspace{5mm} SL Data\\[0.5em]
				\includegraphics[width=0.9\textwidth]{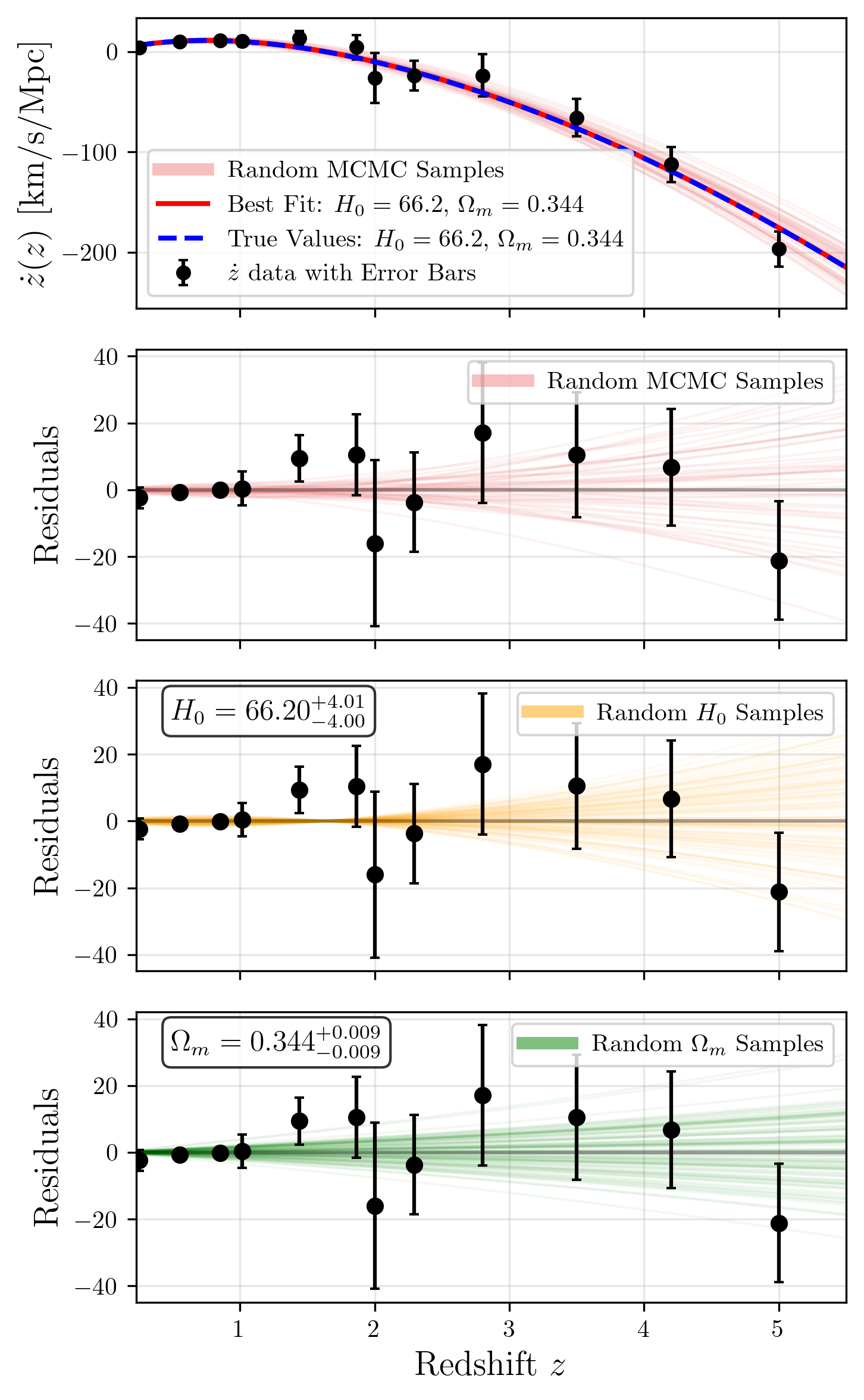} % Reduced width
			\end{minipage}
			\caption{Comparison of fiducial model fitting for CC and simulated SL Data. The top row shows the random MCMC samples (shaded regions), best-fit fiducial models (solid lines), true fiducial values (dashed lines), and simulated data points with error bars. The bottom three rows display residuals between observations and best-fit fiducial models for different parameter constraints: all parameters free (second row), $H_0$ marginalized over other fixed parameters (third row), and $\Omega_m$ marginalized over other fixed parameters (bottom row), with 1$\sigma$ parameter estimates displayed in the upper-left corners. }

			\label{fig:individual_fit}
		\end{figure*}

\begin{figure}[!htbp]
		\centering
		\includegraphics[width=\textwidth]{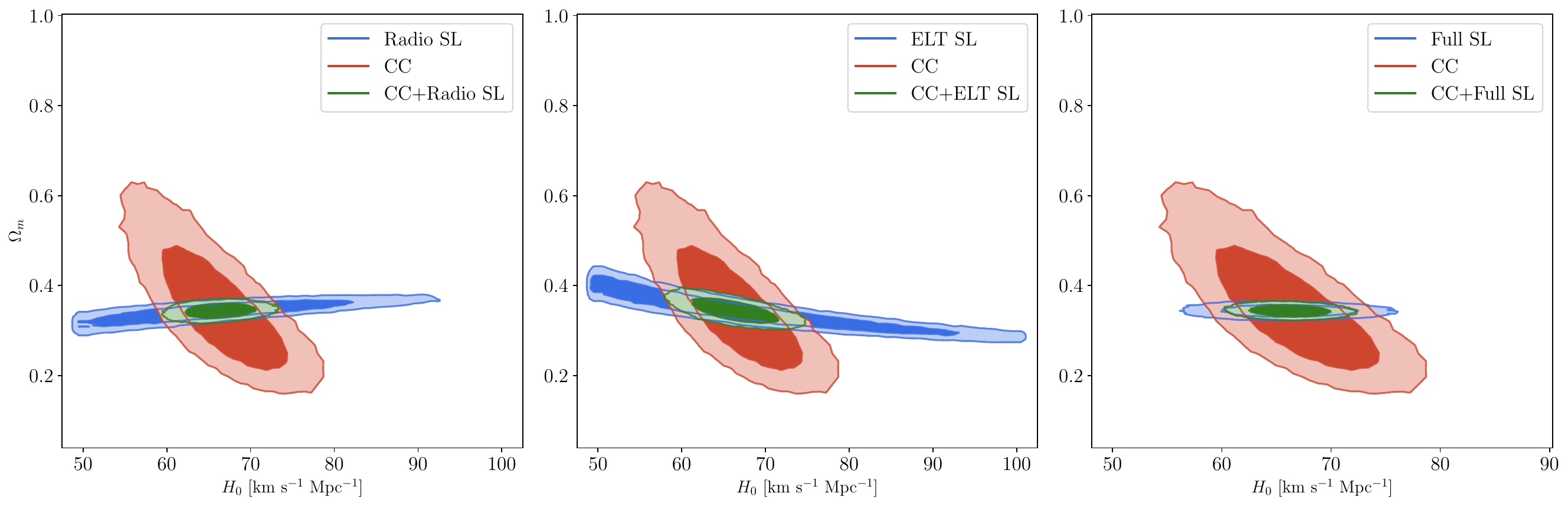}
		\caption{These panels show the 2D joint distributions of the Hubble constant ($H_0$) and matter density parameter ($\Omega_m$), combining CC data with various simulated SL datasets under the assumption of a Flat $\Lambda$CDM model: Radio SL (left panel), ELT SL (middle panel), and Full SL (right panel). Blue, red, and green contours represent constraints from SL, CC, and their joint analysis, respectively. }
		\label{fig:degeneracy}
	\end{figure}

The conventional $\Lambda$CDM parameter fitting approach, while successfully constraining cosmological parameters with reasonable precision, faces inherent limitations from parameter degeneracies and redshift-dependent sensitivity patterns. Figures~\ref{fig:individual_fit} and \ref{fig:degeneracy} illustrate both the strengths and limitations of this methodology when applied to CC and SL datasets.

CC data show increasing MCMC sample dispersion beyond $z \sim 1$, indicating enhanced parameter sensitivity at high redshift, while maintaining tight residual clustering below $z \sim 0.5-1.0$. SL data exhibit a transition around $z \sim 2$, with narrow MCMC dispersion at $z < 2$ becoming notably broader beyond this threshold. Unlike CC, SL maintains consistent observational precision across redshifts, allowing theoretical parameter sensitivity to translate into actual constraining power. The $H_0$ samples show reduced sensitivity near $\dot{z} = 0$, with enhanced constraining power emerging at higher redshifts where the drift signal provides stronger leverage. These patterns yield final parameter uncertainties of $H_0 = 66.22^{+4.87}_{-4.95}$ km s$^{-1}$ Mpc$^{-1}$ and $\Omega_m = 0.344^{+0.103}_{-0.082}$ for CC, versus $H_0 = 66.19^{+4.01}_{-4.00}$ km s$^{-1}$ Mpc$^{-1}$ and $\Omega_m = 0.344^{+0.009}_{-0.009}$ for SL, showing modest $H_0$ improvement but dramatic $\Omega_m$ enhancement.

Figure~\ref{fig:degeneracy} reveals the underlying parameter degeneracy structure that limits individual dataset performance. CC contours (orange) show strong negative $H_0$-$\Omega_m$ correlation with $\sim 45 \degree$ tilted ellipses, arising from parameter coupling in cosmic chronometry. SL constraints (blue) display perpendicular orientation: Radio SL produces broad ellipses nearly orthogonal to CC degeneracy; ELT SL contracts these significantly while preserving orientation; Full SL achieves near-circular contours with balanced precision.

This orthogonal complementarity enables effective degeneracy breaking. Combined CC+SL contours (green) are 1-2 orders of magnitude smaller than individual datasets, approaching circular shapes that eliminate parameter correlations. However, several systematic limitations emerge from this conventional fitting approach: CC residual scatter increases at $z > 1.5$ with outliers reaching $\pm 50$ km s$^{-1}$ Mpc$^{-1}$, reflecting systematic uncertainties in stellar chronometry that compromise high-redshift constraints despite enhanced parameter sensitivity. The progression from Radio to Full SL demonstrates potential for improved precision, but the fundamental challenge remains that parameter degeneracies inherent to individual probes limit their standalone constraining power, necessitating multi-probe approaches for optimal cosmological inference. 

	\begin{figure*}[htbp]
		\centering
		\begin{minipage}[b]{0.45\textwidth}
			\centering
			\hspace{5mm} CC Data\\[0.5em]
			\includegraphics[width=0.9\textwidth]{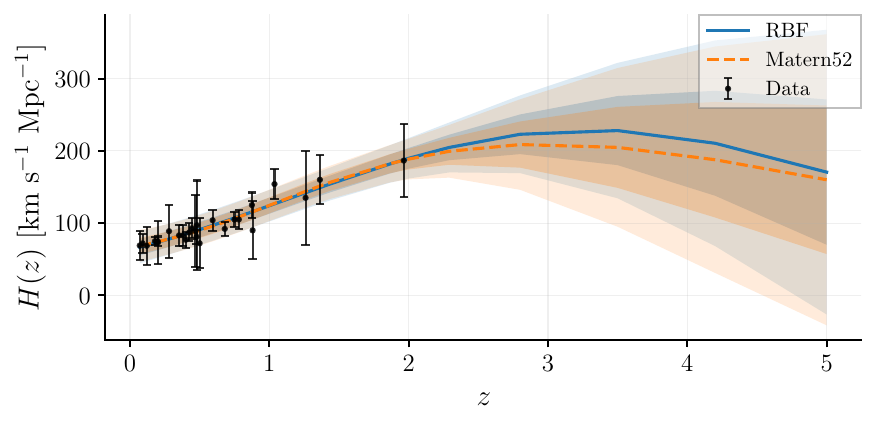} % Reduced width
		\end{minipage}
		\hfill
		\begin{minipage}[b]{0.45\textwidth}
			\centering
			\hspace{5mm} SL Data\\[0.5em]
			\includegraphics[width=0.9\textwidth]{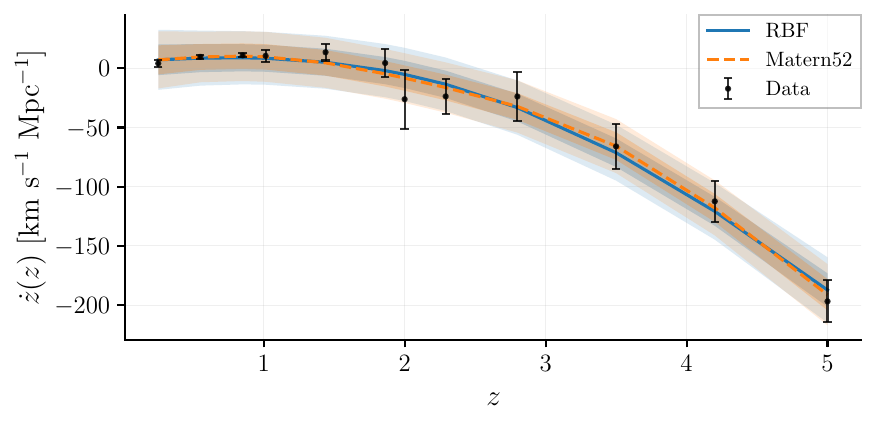} % Reduced width
		\end{minipage}
		\caption{GP  reconstructions of CC (left) and SL (right) data utilizing RBF (blue) and Matérn 5/2 (orange) kernels. The shaded regions represent the 68\% (darker) and 95\% (lighter) confidence intervals, illustrating the uncertainty in the GP fits. }

		\label{fig:gp_reconstruction}
	\end{figure*}

GP regression provides a standard model-independent approach for reconstructing cosmological functions and aligning disparate datasets in redshift \cite{rasmussen2006gaussian, 2012JCAP...06..036S}. We implement this baseline methodology using the GPy framework, selecting the widely adopted RBF and Matérn 5/2 kernels for their different smoothness assumptions. To ensure a robust reconstruction and mitigate optimizer-induced systematics, we rigorously determine the kernel hyperparameters by maximizing the marginal log-likelihood via Bayesian optimization (30 iterations). The parameter bounds are constrained to physically motivated ranges: $\sigma_f^2 \in [10^{-3}, 10^{2}]$ for the signal variance and $\ell \in [0.1, 5.0]$ for the characteristic correlation length. This careful procedure is essential to minimize bias in the reconstructed $H(z)$ and its extrapolation to $z=0$ for $H_0$ estimation. Finally, the reconstruction uncertainties and the error on $H_0$ are derived analytically from the posterior predictive distribution, providing the exact conditional uncertainty for our Gaussian likelihood model.
Figure~\ref{fig:gp_reconstruction} shows GP reconstructions using RBF (blue) and Matérn 5/2 (orange) kernels, serving as a direct comparison baseline for the Geometric Embedding method developed in this work. Both kernels produce nearly identical reconstructions across the full redshift range. For CC data (left), the smooth monotonic increase with redshift captures the expected Hubble parameter evolution at low-to-intermediate redshifts, with confidence intervals (68\% and 95\% shaded regions) reflecting data density and precision. However, beyond $z \sim 1$ where data become sparse and noisy, the GP reconstruction exhibits large uncertainties and an unphysical declining trend in regions lacking observational constraints. This degradation reflects the fundamental limitation of GP methods when extrapolating beyond dense data coverage. The extrapolation to $z=0$ for $H_0$ estimation further illustrates this challenge: RBF kernel yields $H_0 = 63.18 \pm 11.67$~km\,s$^{-1}$\,Mpc$^{-1}$ while Matérn 5/2 produces $H_0 = 65.45 \pm 11.73$~km\,s$^{-1}$\,Mpc$^{-1}$, demonstrating both kernel-dependent systematics and prohibitively large uncertainties ($\sim 18\%-19\%$) that severely limit the practical utility of GP extrapolation for precision cosmology.

SL data (right) demonstrate the complex evolution of $\dot{z}$, exhibiting an initial increase followed by a decline that crosses zero, reflecting the transition from deceleration to acceleration epochs in cosmic expansion history, with more consistent behavior across the full redshift range due to better data coverage. The minimal differences between kernels validate reconstruction robustness and confirm that kernel choice does not significantly impact recovered cosmic evolution. Given this equivalence, we adopt the Matérn 5/2 kernel for its superior smoothness properties and finite differentiability characteristics that better capture expected cosmological behavior. These GP reconstructions highlight the challenges of conventional non-parametric methods in handling sparse, noisy data, providing a benchmark against which the performance and advantages of our Geometric Embedding approach can be directly assessed.

	\begin{figure}[hbp]
		\centering
		\includegraphics[width=\textwidth]{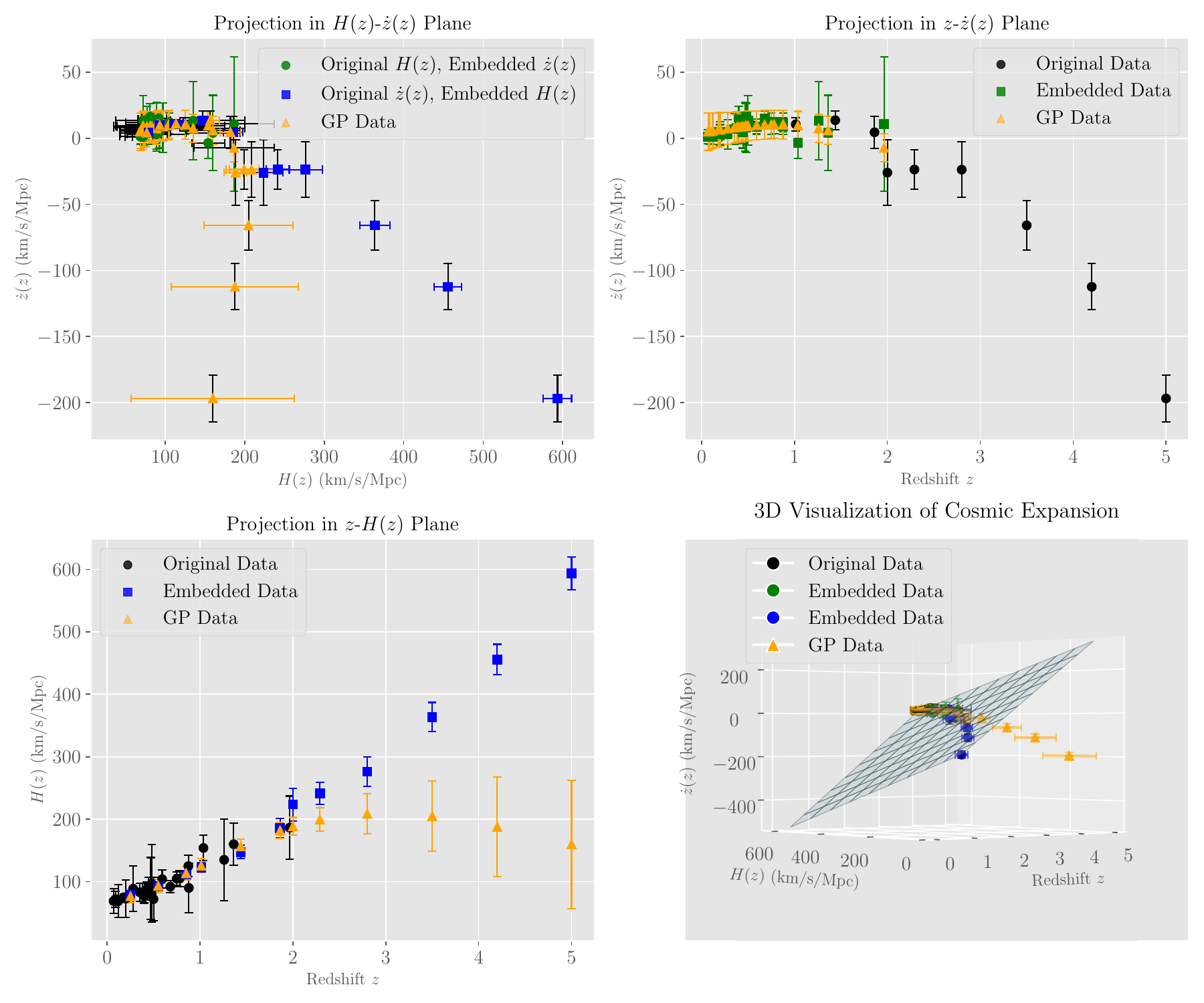}
		    \caption{Visualization of the geometric embedding method in $(z, H(z), \dot{z})$ space with projections in the $H(z)$-$\dot{z}$ (top left), $z$-$\dot{z}$ (top right), $z$-$H(z)$ (bottom left) planes and a 3D view (bottom right). Black error bars represent original measurements. Green and blue error bars show geometric embedding results for $\dot{z}$ and $H(z)$ respectively. Orange error bars display GP reconstruction. The figure demonstrates how our method transforms redshift misalignment into a geometric constraint without requiring interpolation.}

		\label{fig:geometric_embedding}
	\end{figure}

Before proceeding to detailed quantitative comparisons, Figure~\ref{fig:geometric_embedding} visualizes the practical implementation of our geometric embedding approach with full data. The projections in different planes reveal the relationship between observed quantities and their embedded counterparts, showing how our method maintains consistency with both observational data and kinematic relations. This visual representation makes concrete the abstract concept of a geometric plane in $(z, H(z), \dot{z})$ space and demonstrates the difference between our physically-motivated approach and the statistically-driven GP method. The 3D perspective particularly highlights how embedded points naturally conform to the kinematic surface regardless of their original redshift positions.

	\begin{figure}[htbp]
		\centering
		
		% 列标题 - 普通字号，居中
		\begin{minipage}[t]{0.45\textwidth}
			\centering
			\hspace{16mm} Geometric Embedding
		\end{minipage}
		\hfill
		\begin{minipage}[t]{0.45\textwidth}
			\centering
			\hspace{5mm} GP Reconstruction
		\end{minipage}
		
		\vspace{0.25em}
		
		% 第一行
		\begin{minipage}[c]{0.05\textwidth}
			\centering
			\hspace{5mm}\rotatebox{90}{CC + Radio SL}
		\end{minipage}
		\begin{minipage}[c]{0.45\textwidth}
			\centering
			\includegraphics[width=\textwidth,height=5.8cm]{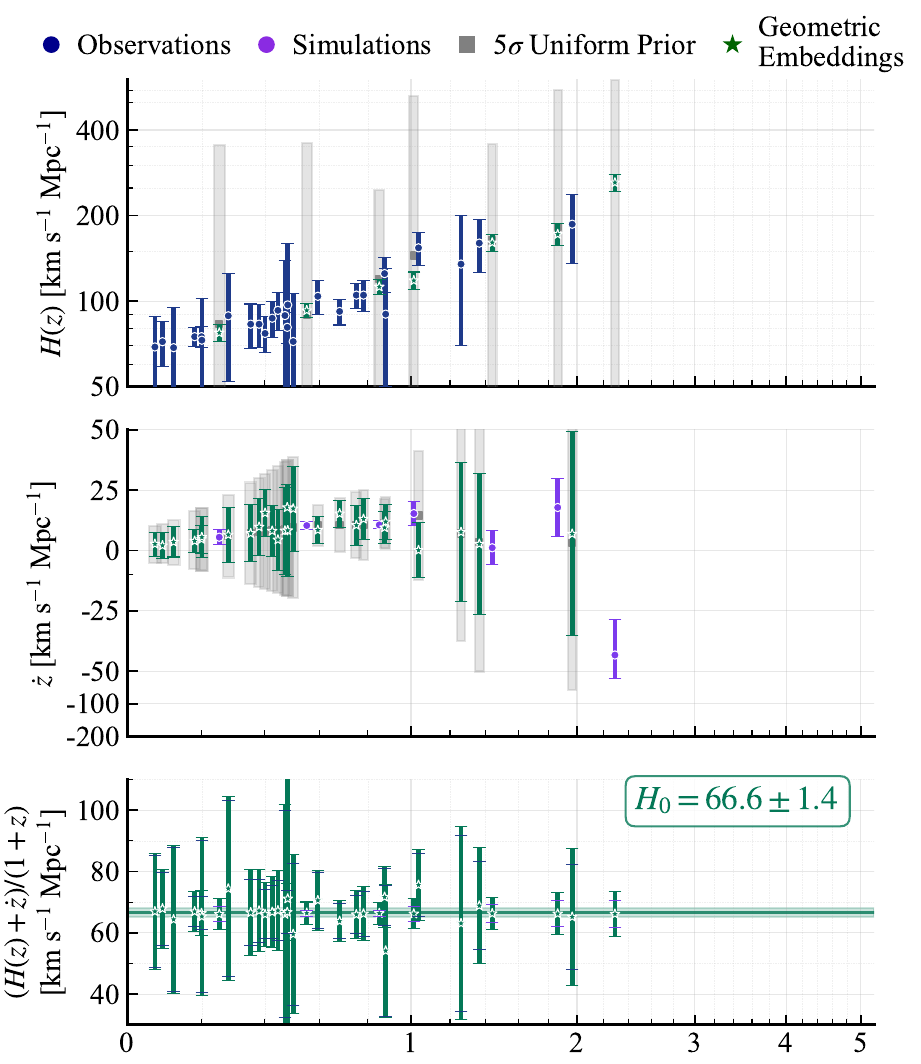}
		\end{minipage}
		\hfill
		\begin{minipage}[c]{0.45\textwidth}
			\centering
			\includegraphics[width=\textwidth,height=5.8cm ]{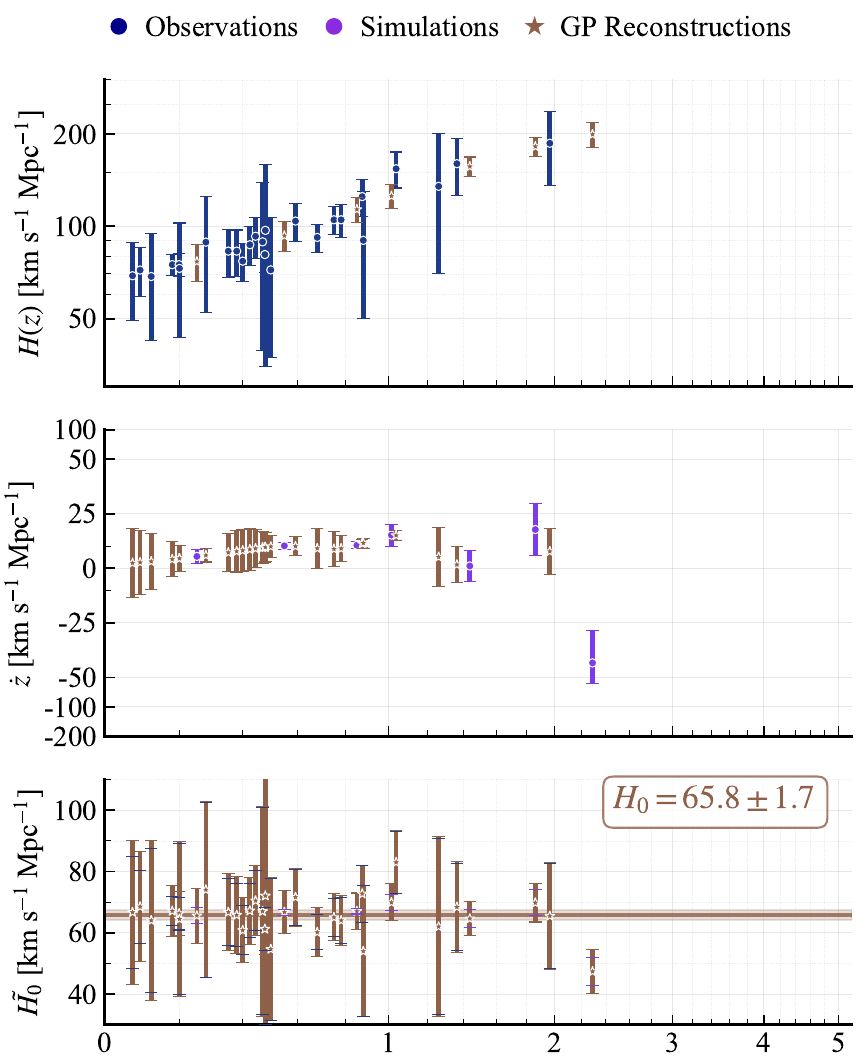}
		\end{minipage}

		% 第二行
		\begin{minipage}[c]{0.05\textwidth}
			\centering
			\rotatebox{90}{CC + ELT SL}
		\end{minipage}
		\begin{minipage}[c]{0.45\textwidth}
			\centering
			\includegraphics[width=\textwidth,height=5.8cm]{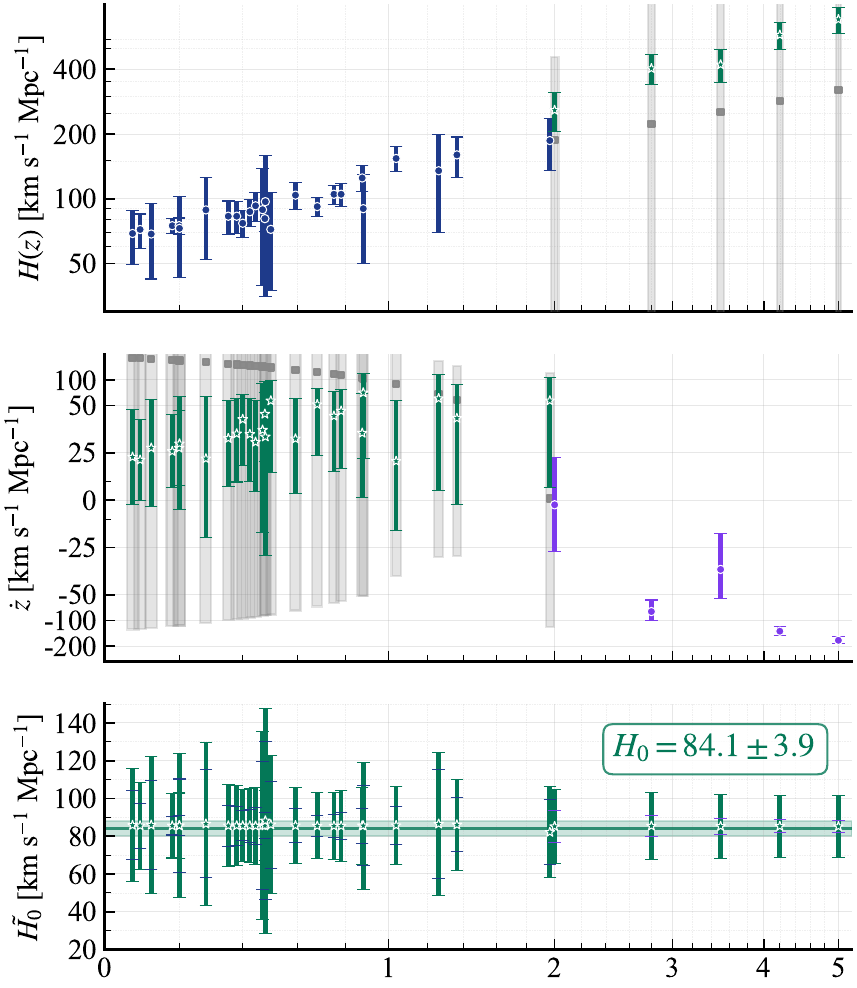}
		\end{minipage}
		\hfill
		\begin{minipage}[c]{0.45\textwidth}
			\centering
			\includegraphics[width=\textwidth,height=5.8cm ]{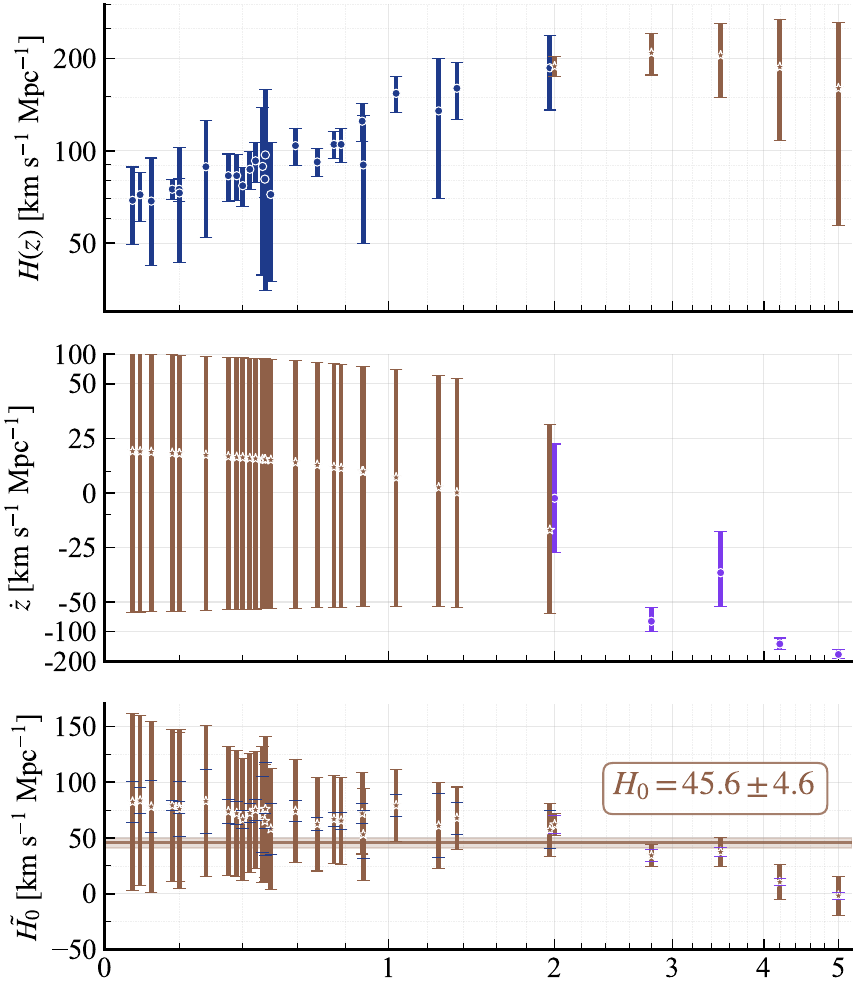}
		\end{minipage}

		% 第三行
		\begin{minipage}[c]{0.05\textwidth}
			\centering
			\rotatebox{90}{CC + Full SL}
		\end{minipage}
		\begin{minipage}[c]{0.45\textwidth}
			\centering
			\includegraphics[width=\textwidth,height=5.8cm ]{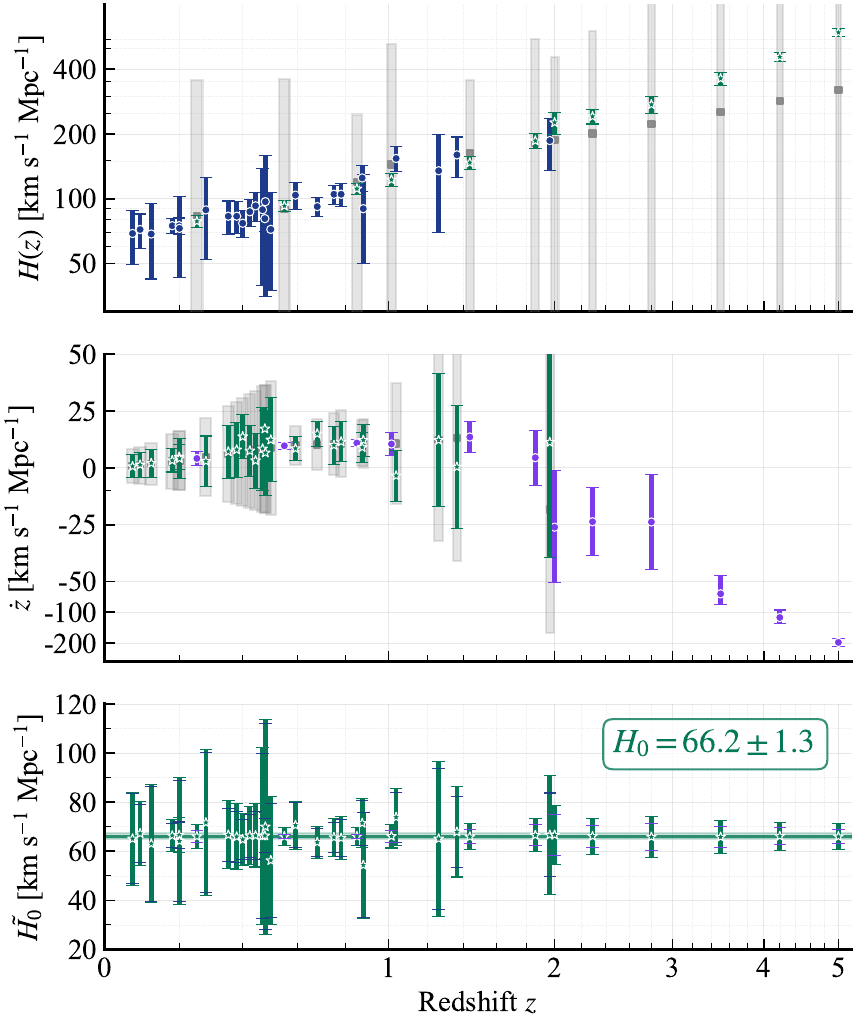}
		\end{minipage}
		\hfill
		\begin{minipage}[c]{0.45\textwidth}
			\centering
			\includegraphics[width=\textwidth,height=5.8cm ]{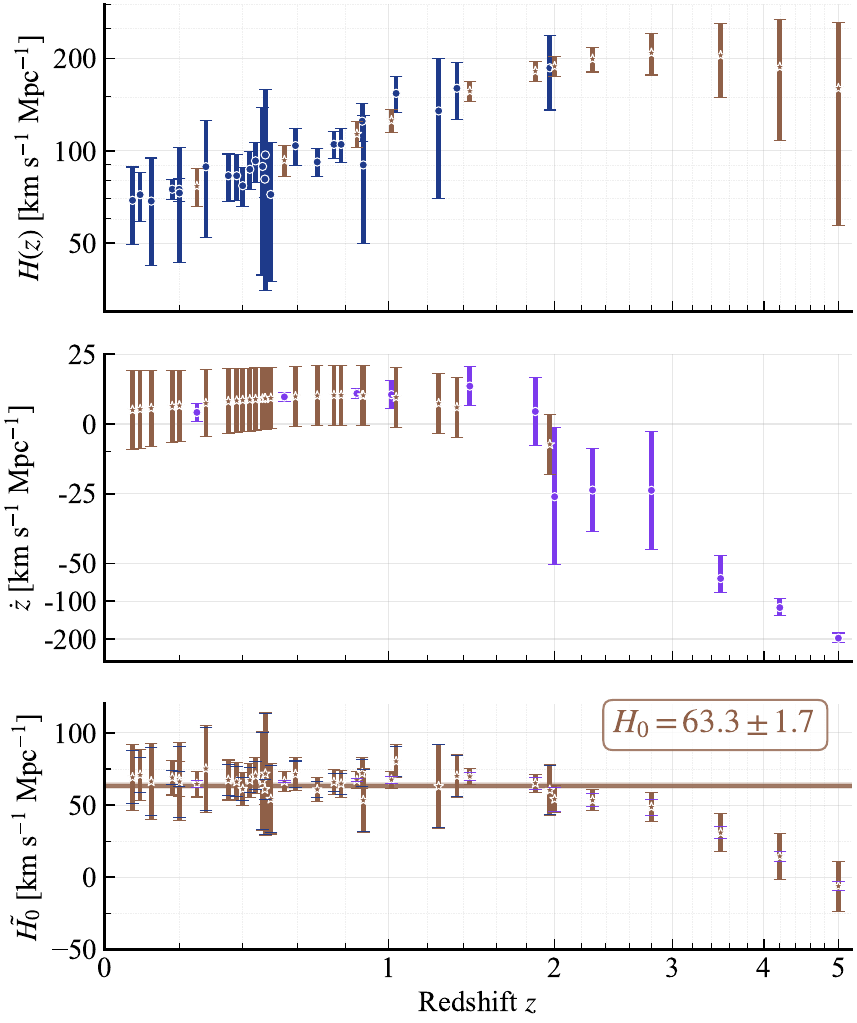}
		\end{minipage}
		
		\caption{Comparison between Geometric Embedding (left) and GP Reconstruction (right) methods across CC + Radio SL (top), CC + ELT SL (middle), and CC + Full SL (bottom) data combinations. Each row displays $H(z)$, $\dot{z}$, and derived $H_{0}$ estimates. Grey regions show uniform priors centered on grey squares; blue/purple points represent observations/simulations with error bars. Green/brown asterisks indicate reconstructed values. Bottom panels show individual $\tilde{H}_0$ estimates (error bars) with inverse variance weighted means (colored bands).}
		\label{fig:geometric_embedding_results}
	\end{figure}

Figure~\ref{fig:geometric_embedding_results} provides a detailed visual comparison of redshift alignment performance between Geometric Embedding and GP methods across three data combinations with distinctly different redshift overlap characteristics. When redshift coverage overlaps substantially, both alignment methods perform remarkably well. CC+Radio SL exemplifies this ideal scenario, where CC data extending to $z \lesssim 2$ connects smoothly with Radio SL observations. Both approaches produce clean, monotonically increasing $H(z)$ reconstructions and capture the expected deceleration-acceleration transition in $\dot{z}$ around $z \sim 0.5-1$. Most importantly, their $H_0$ distributions appear well-behaved and Gaussian, centered around consistent values. This visual concordance demonstrates that sufficient redshift overlap enables reliable reconstruction regardless of the specific alignment technique.

Complete redshift disconnection presents a fundamentally different challenge. CC+ELT SL forces both methods to bridge a substantial gap between low-redshift CC data ($z \lesssim 2$) and isolated high-redshift ELT points around $z \sim 2-5$. While individual $H(z)$ reconstructions appear reasonable within data-rich regions, extrapolation across this desert triggers catastrophic failures. GP reconstruction collapses to an anomalously tight, low-$H_0$ distribution, whereas Geometric Embedding swings to the opposite extreme with a broad, elevated distribution. Even the $\dot{z}$ reconstructions become problematic, with GP showing volatile oscillations and Geometric Embedding displaying smoother but potentially biased transitions.

Partial overlap scenarios reveal crucial differences in method robustness. CC+Full SL combines the dense CC baseline with SL points that partially fill intermediate redshifts, yet still leaves gaps at the highest redshifts where ELT data resides. Under these mixed conditions, GP reconstruction shows clear contamination effects—its $H_0$ distribution becomes systematically biased and skewed despite the presence of some connecting data. Geometric Embedding, however, maintains remarkable stability, producing a well-centered, symmetric $H_0$ distribution that appears largely unaffected by the problematic high-redshift component.

Error propagation patterns further illuminate these methodological differences. Overlapping coverage keeps uncertainties well-controlled for both approaches. Non-overlapping cases inflate errors dramatically, though with distinct signatures—GP tends toward more erratic uncertainty estimates while Geometric Embedding shows more systematic expansion. Under partial overlap, Geometric Embedding demonstrates superior error management, maintaining stable uncertainty propagation despite the challenging redshift structure that clearly compromises GP performance. These visual diagnostics underscore how redshift coverage quality determines alignment method success, with Geometric Embedding showing notably better resilience to the coverage gaps that severely compromise GP-based approaches.

	\begin{table}[!htbp]
		\centering
		\caption{Hubble constant measurements from different analysis methods}
		\label{tab:h0_results}
		\begin{tabular}{@{}lcccccc@{}}
		\hline
		\hline
		Method & Data Size & $H_0$ & Uncertainty & Precision & Offset $^{\rm a}$ & \\
		& & (km\,s$^{-1}$\,Mpc$^{-1}$) & (km\,s$^{-1}$\,Mpc$^{-1}$) & (\%) & ($\sigma$) & \\
		\hline  \\
		\multicolumn{6}{c}{\textit{Individual $\Lambda$CDM Fitting} } \\
		% CC-only ($\Lambda$CDM) & 68.85 & $\pm$4.10 & 6.0 \\
		CC & 26 & 66.17 & $\pm$4.89 & 7.4 & 0 & \\
		Radio SL & 7 & 65.90 & $\pm$10.20 & 15.5 & -0.03 & \\
		ELT SL & 5 & 65.87 & $\pm$13.67 & 20.7 &  -0.02& \\
		Full SL & 12 & 66.19 & $\pm$4.00 & 6.0 & 0.005 & \\ 
		\hline \\
		\multicolumn{6}{c}{\textit{Joint analyses}} \\
		\hline \\
		\multicolumn{6}{c}{\textit{Joint $\Lambda$CDM Fitting}} \\
		CC + Radio SL & 26 + 7 & 66.19 & $\pm$2.82  & 4.3  &  0.01 & \\
		CC + ELT SL & 26 + 5 & 66.20 & $\pm$ 3.43 & 5.2  &  0.01 & \\
		CC + Full SL & 26 + 12 & 66.30 & $\pm$2.42 & 3.7 & 0.05 & \\ 
		\hline \\
		\multicolumn{6}{c}{\textit{GP Reconstructions}} \\
		CC extrap. (RBF) & 26 & 63.18 & $\pm$11.67 & 18.5 & -0.26 & \\
		CC extrap. (Matérn 5/2)& 26 & 65.45 & $\pm$11.73 & 17.9 & -0.61 & \\
		CC + Radio SL & 26 + 7 & 65.83 & $\pm$1.68 & 2.6 & -0.20 & \\
		CC + ELT SL & 26 + 5 & 45.64 & $\pm$4.63 & 10.1 & +4.43 & \\
		CC + Full SL & 26 + 12 & 63.27 & $\pm$1.71 & 1.9 & -1.71 & \\ 
		\hline \\
		\multicolumn{6}{c}{\textit{Geometric Embedding}} \\
		CC + Radio SL & 26 + 7 & 66.59 & $\pm$1.41 & 2.1 & 0.30 & \\
		CC + ELT SL & 26 + 5 & 84.14 & $\pm$3.92 & 4.6 & +4.58 & \\
		CC + Full SL & 26 + 12 & 66.26& $\pm$1.26 & 1.9 & 0.02 & \\
		\hline \\
		\multicolumn{6}{c}{\textcolor{black}{\textit{CC + Full SL Geometric Embedding: Prior Sensitivity}}} \\
		\textcolor{black}{$\mathcal{U}(\mu \pm 1\sigma)$} & \textcolor{black}{26 + 12} & \textcolor{black}{65.05} & \textcolor{black}{$\pm$1.12} & \textcolor{black}{1.7} & \textcolor{black}{-0.22} & \\
		\textcolor{black}{$\mathcal{U}(\mu \pm 3\sigma)$} & \textcolor{black}{26 + 12} & \textcolor{black}{65.96} & \textcolor{black}{$\pm$1.19} & \textcolor{black}{1.8} & \textcolor{black}{-0.04} & \\
		\textcolor{black}{$\mathcal{U}(\mu \pm 7\sigma)$} & \textcolor{black}{26 + 12} & \textcolor{black}{66.62} & \textcolor{black}{$\pm$1.36} & \textcolor{black}{2.0} & \textcolor{black}{0.09} & \\
		\hline
		\hline
		\multicolumn{6}{@{}l@{}}{\footnotesize $^{\rm a}$ Offset from the fiducial $H_0$ value} \\
		\end{tabular}
		% \label{tab:h0_results}
	\end{table}

Table~\ref{tab:h0_results} presents a comprehensive comparison of $H_0$ measurements across different analysis approaches and data combinations. The results reveal several important insights about the relative performance of these methodologies. Joint analyses consistently outperform individual dataset analysis in terms of precision. The improvement is substantial: while CC data alone yields $H_0 = 66.17 \pm 4.89$~km\,s$^{-1}$\,Mpc$^{-1}$ (7.4\% uncertainty), joint approaches significantly reduce uncertainties. For example, joint $\Lambda$CDM fitting of CC with Radio SL data achieves $\pm 2.82$~km\,s$^{-1}$\,Mpc$^{-1}$ (4.3\%), while Geometric Embedding with the same data combination reaches even higher precision at $\pm 1.41$~km\,s$^{-1}$\,Mpc$^{-1}$ (2.1\%). This clearly demonstrates the statistical advantages of multi-probe analysis.

The robustness of the Geometric Embedding methodology is further validated through the sensitivity analysis of prior width choices shown in the lower section of Table~\ref{tab:h0_results}. This analysis was conducted to specifically address the influence of prior selection, demonstrating that the inferred Hubble constant is not sensitive to the exact choice of boundary. The results form a clear plateau of stability: across a wide range of priors ($\pm1\sigma$ to $\pm7\sigma$), the statistical offset from the fiducial value remains minimal and stable, confined within a narrow range of $\pm0.3\sigma$. Concurrently, the measurement precision remains consistently high ($1.7\%$--$2.0\%$). This confirms that our fiducial $\pm5\sigma$ prior—initially chosen to be conservatively wide—resides well within this stable region and does not dictate the scientific conclusion. The stability of both the offset and precision underscores that the results are robust and driven by the constraints of the data and the kinematic model, rather than by prior assumptions.

However, the redshift alignment methods face significant challenges when dealing with datasets that have minimal redshift overlap. The CC+ELT SL combination proves particularly problematic for both GP reconstruction and Geometric Embedding. GP yields an unrealistically low $H_0 = 45.64 \pm 4.63$~km\,s$^{-1}$\,Mpc$^{-1}$ (+4.43$\sigma$ deviation), while Geometric Embedding produces an anomalously high $H_0 = 84.14 \pm 3.92$~km\,s$^{-1}$\,Mpc$^{-1}$ (+4.58$\sigma$ deviation). This systematic failure highlights the fundamental difficulty of extrapolating across disconnected redshift ranges. It demonstrates that even physics-based geometric constraints cannot extract information where none exists; the kinematic relation $\dot{z} = H_0(1+z) - H(z)$ becomes mathematically underdetermined without overlapping data to anchor the solution, a fundamental limitation shared by any model-independent approach.

Interestingly, the GP reconstruction is more susceptible to these non-overlap issues: even for CC+Full SL where ELT data comprises only a portion of the sample, GP still yields $H_0 = 63.27 \pm 1.71$~km\,s$^{-1}$\,Mpc$^{-1}$ with a notable -1.71$\sigma$ offset. In contrast, Geometric Embedding with the same CC+Full SL combination shows remarkable robustness, achieving $H_0 = 66.26 \pm 1.26$~km\,s$^{-1}$\,Mpc$^{-1}$ with minimal bias (0.02$\sigma$ offset). This stark difference underscores a key advantage: our method's reliance on fundamental kinematic constraints provides inherent resilience against the biases that affect methods relying on statistical smoothness priors (GP), as long as some degree of redshift connectivity is present.

Remarkably, when we exclude the problematic non-overlapping combinations, all other $H_0$ measurements show excellent consistency with deviations well below 1$\sigma$. Most offsets remain within $\pm 0.3\sigma$, indicating good systematic control across different approaches and data combinations despite methodological differences.

Among the redshift alignment approaches, Geometric Embedding emerges as the clear winner. For all reliable data combinations, it delivers the highest precision measurements while maintaining minimal systematic biases. This superior performance likely stems from the method's geometric consistency constraints, which provide better resilience against the extrapolation pitfalls that challenge traditional GP approaches. The anomalous result for the CC+ELT SL combination thus serves not to limit the method's applicability, but to define its domain of validity and act as a built-in diagnostic for data connectivity. The robust performance in the CC+Full SL case confirms its practical utility for realistic future data combinations that feature the necessary degree of redshift overlap.

\section{Discussions}
\label{sec:discussions}
Prior model-independent CC reconstructions suggested the Hubble tension may not be significant, but their large uncertainties ($\sim$7-15\%) provided insufficient evidence for definitive conclusions. The precision achieved through geometric embedding enables $H_0$ constraints at the few-percent level without relying on early-universe physics or local calibration hierarchies, providing the statistical power necessary to definitively assess whether the Hubble tension reflects systematic uncertainties or genuine departures from $\Lambda$CDM cosmology.

The geometric embedding approach achieves superior precision through global consistency requirements that naturally resolve redshift misalignment between CC and SL datasets without functional assumptions about redshift evolution. This kinematic coupling enables cross-validation and mutual error suppression between independent systematic uncertainties of different observational channels. Additionally, the approach mitigates parameter degeneracies by reducing the system to a single true degree of freedom: $H_0$. While formally introducing $N$ embedded quantities, the fundamental kinematic constraint links all observations to this underlying parameter, creating a physics-informed regularization where direct algebraic constraints extract information more efficiently than parametric fitting—accounting for the $\sim$30\% precision improvement over joint $\Lambda$CDM fitting.

Our embedding priors, derived through interpolation/extrapolation from adjacent observations, establish 5$\sigma$ bounds with uniform distributions representing minimal statistical assumptions. As demonstrated in Figure~\ref{fig:geometric_embedding_results}, posterior distributions are significantly more compact than these prior bounds, indicating that kinematic constraints—rather than imposed boundaries—primarily determine parameter estimates. This differs fundamentally from GP methods that impose prescribed correlation structures, maintaining model independence while providing robust statistical inference through data-driven regularization.

The methodology remains constrained by the FLRW assumption, limiting applicability across different cosmological scenarios. Most theoretical frameworks addressing the Hubble tension preserve the cosmological principle while altering expansion dynamics, making them compatible with our kinematic approach. More exotic scenarios such as local inhomogeneity models or anisotropic spacetimes formally violate FLRW assumptions but show minor deviations within our observational window ($z \lesssim 5$). Beyond $H_0$ determination, our approach provides model-independent data reconstruction for cosmological constraints and offers a direct observational test for potential FLRW violations through systematic deviations in the reconstructed $H(z)$.

Regarding noise pattern selection during validation, this strategy assessed method robustness under realistic conditions while avoiding numerical artifacts. The noise-free validation yielded consistent $H_0$ constraints ($66.72 \pm 1.25$ km s$^{-1}$ Mpc$^{-1}$), confirming stable performance. The observed offset from the fiducial input ($66.17$ km s$^{-1}$ Mpc$^{-1}$) reflects expected nonlinear estimation effects where Jensen's inequality guarantees bias proportional to noise level and estimator curvature—the $\sim 0.8\%$ offset falls within expected ranges for our simulated conditions.

The ability to measure $H_0$ independently at each observed redshift provides significant advantages for investigating effective running Hubble constant scenarios compared to traditional redshift binning approaches. Conventional studies divide data into discrete bins and assume constant $H_0$ within each bin, introducing systematic uncertainties through arbitrary boundary choices and averaging effects that smooth out genuine variations. Our geometric embedding approach extracts individual $\tilde{H}_{0,i}$ values directly at each observed redshift without binning assumptions, providing continuous coverage that can detect subtle evolution patterns—including non-monotonic variations—while maintaining full statistical power of the observational dataset for constraining potential running $H_0(z)$ .

While our geometric embedding approach already demonstrates robust and precise $H_0$ constraints using current CC and prospective SL datasets with promising precision, the method's full potential will be realized through enhanced high-redshift CC observations from upcoming facilities. Large-scale spectroscopic surveys like Dark Energy Spectroscopic Instrument (DESI) and 4-metre Multi-Object Spectroscopic Telescope (4MOST) will significantly expand reliable CC measurements to higher redshifts, directly addressing the current limitations of sparse, noisy, and non-overlapping datasets that challenge conventional methods, while space-based capabilities such as JWST and the Roman Space Telescope will extend stellar age determination capabilities, providing crucial high-redshift CC data points that fill critical gaps in redshift coverage. The anticipated denser redshift sampling will particularly benefit our geometric embedding framework by mitigating the sparse data problem that currently limits precision, reducing observational noise through increased statistics, and establishing better redshift overlap between CC and SL components—improvements that directly address the core challenges that make GP methods systematically biased under disconnected coverage scenarios, while our physics-based kinematic constraints maintain robustness. Next-generation facilities including ELT-ANDES will further enable SL redshift drift measurements through improved spectroscopic precision, providing complementary kinematic constraints that enhance our $\tilde{H}_{0,i}$ extraction capabilities and strengthen the overall geometric embedding approach for model-independent $H_0$ determination.

\section{Conclusion}
\label{sec:conclusion}
In this article, we demonstrate that a data-driven, physics-induced method for high-precision $H_0$ measurement is achievable at the 2.1\% level while maintaining complete model independence. Our approach leverages the fundamental kinematic relation $\dot{z} = H_0(1+z) - H(z)$ to transform the observational alignment challenge between CC and SL datasets into a geometric constraint problem. We show that $H_0$ directly determines the normal vector of the FLRW observational plane, establishing mathematical constraints under minimal assumptions.

The central finding lies in recognizing that apparent observational misalignment between CC and SL datasets can be exploited as physics-informed regularization rather than treated as a technical obstacle. Using CC and SL, our method transforms this geometric relationship into direct algebraic determination of the current expansion rate, achieving substantial precision improvements over conventional approaches---demonstrating $\sim$30\% enhancement compared to joint $\Lambda$CDM fitting and superior robustness against the extrapolation failures that severely limit GP methods.

Comprehensive validation reveals critical distinctions in methodological performance under varying redshift coverage scenarios. While both geometric embedding and GP reconstruction succeed under substantial redshift overlap, disconnected datasets expose fundamental differences: GP methods become systematically biased even with partial coverage gaps, whereas geometric embedding maintains statistical consistency through its physics-based constraint structure. This resilience stems from the method's reliance on kinematic relationships rather than imposed smoothness assumptions.

Our results establish that high-precision, model-independent cosmology is achievable through geometric methods that harness fundamental physical constraints, yielding robust model-independent estimates. The approach yields $H_0 = 66.26 \pm 1.26$ km s$^{-1}$ Mpc$^{-1}$ (1.9\% precision) for well-aligned datasets. This geometric framework establishes a fundamentally new approach to cosmological measurements through pure observational analysis, opening new avenues for precision cosmology through kinematic constraints. The methodological validation presented here provides a robust analytical framework that awaits implementation with the first generation of actual redshift drift measurements, connecting contemporary theoretical foundations with future observational capabilities.

\section*{Acknowledgments}

We thank the anonymous referees for their constructive comments and suggestions that significantly improved the clarity and presentation of this manuscript. We also thank Rui Li for helpful discussions and support.
This work was supported by the National Natural Science Foundation of China (Nos.\ 12403004, 12588202 and 12473002), the national Key Program for Science and Technology Research Development (Nos. 2023YFB3002500 and  2024YFA1611804),  the China Manned Space Program with grant No. CMS-CSST-2025-A01, and the Postdoctoral Fellowship Program (Grade C) of  China Postdoctoral Science Foundation(GZC20241563).

\bibliographystyle{elsarticle-num-names}
\bibliography{mybib} 

\begin{thebibliography}{47}
\expandafter\ifx\csname natexlab\endcsname\relax\def\natexlab#1{#1}\fi
\providecommand{\url}[1]{\texttt{#1}}
\providecommand{\href}[2]{#2}
\providecommand{\path}[1]{#1}
\providecommand{\DOIprefix}{doi:}
\providecommand{\ArXivprefix}{arXiv:}
\providecommand{\URLprefix}{URL: }
\providecommand{\Pubmedprefix}{pmid:}
\providecommand{\doi}[1]{\href{http://dx.doi.org/#1}{\path{#1}}}
\providecommand{\Pubmed}[1]{\href{pmid:#1}{\path{#1}}}
\providecommand{\bibinfo}[2]{#2}
\ifx\xfnm\relax \def\xfnm[#1]{\unskip,\space#1}\fi
%Type = Article
\bibitem[{{Planck Collaboration}(2020)}]{2020A&A...641A...6P}
\bibinfo{author}{{Planck Collaboration}},
\newblock \bibinfo{title}{{Planck 2018 results. VI. Cosmological parameters}},
\newblock \bibinfo{journal}{\aap} \bibinfo{volume}{641} (\bibinfo{year}{2020})
  \bibinfo{pages}{A6}. \DOIprefix\doi{10.1051/0004-6361/201833910}.
  \href{http://arxiv.org/abs/1807.06209}{{\tt arXiv:1807.06209}}.
%Type = Article
\bibitem[{{Riess} et~al.(2022){Riess}, {Yuan}, {Macri}, {Scolnic}, {Brout},
  {Casertano}, {Jones}, {Murakami}, {Anand}, {Breuval}, {Brink}, {Filippenko},
  {Hoffmann}, {Jha}, {D'arcy Kenworthy}, {Mackenty}, {Stahl}, and
  {Zheng}}]{2022ApJ...934L...7R}
\bibinfo{author}{A.~G. {Riess}}, \bibinfo{author}{W.~{Yuan}},
  \bibinfo{author}{L.~M. {Macri}}, \bibinfo{author}{D.~{Scolnic}},
  \bibinfo{author}{D.~{Brout}}, \bibinfo{author}{S.~{Casertano}},
  \bibinfo{author}{D.~O. {Jones}}, \bibinfo{author}{Y.~{Murakami}},
  \bibinfo{author}{G.~S. {Anand}}, \bibinfo{author}{L.~{Breuval}},
  \bibinfo{author}{T.~G. {Brink}}, \bibinfo{author}{A.~V. {Filippenko}},
  \bibinfo{author}{S.~{Hoffmann}}, \bibinfo{author}{S.~W. {Jha}},
  \bibinfo{author}{W.~{D'arcy Kenworthy}}, \bibinfo{author}{J.~{Mackenty}},
  \bibinfo{author}{B.~E. {Stahl}}, \bibinfo{author}{W.~{Zheng}},
\newblock \bibinfo{title}{{A Comprehensive Measurement of the Local Value of
  the Hubble Constant with 1 km s$^{-1}$ Mpc$^{-1}$ Uncertainty from the Hubble
  Space Telescope and the SH0ES Team}},
\newblock \bibinfo{journal}{\apjl} \bibinfo{volume}{934} (\bibinfo{year}{2022})
  \bibinfo{pages}{L7}. \DOIprefix\doi{10.3847/2041-8213/ac5c5b}.
  \href{http://arxiv.org/abs/2112.04510}{{\tt arXiv:2112.04510}}.
%Type = Article
\bibitem[{{Wong} et~al.(2020){Wong}, {Suyu}, {Chen}, {Rusu}, {Millon}, {Sluse},
  {Bonvin}, {Fassnacht}, {Taubenberger}, {Auger}, {Birrer}, {Chan}, {Courbin},
  {Hilbert}, {Tihhonova}, {Treu}, {Agnello}, {Ding}, {Jee}, {Komatsu},
  {Shajib}, {Sonnenfeld}, {Blandford}, {Koopmans}, {Marshall}, and
  {Meylan}}]{2020MNRAS.498.1420W}
\bibinfo{author}{K.~C. {Wong}}, \bibinfo{author}{S.~H. {Suyu}},
  \bibinfo{author}{G.~C.~F. {Chen}}, \bibinfo{author}{C.~E. {Rusu}},
  \bibinfo{author}{M.~{Millon}}, \bibinfo{author}{D.~{Sluse}},
  \bibinfo{author}{V.~{Bonvin}}, \bibinfo{author}{C.~D. {Fassnacht}},
  \bibinfo{author}{S.~{Taubenberger}}, \bibinfo{author}{M.~W. {Auger}},
  \bibinfo{author}{S.~{Birrer}}, \bibinfo{author}{J.~H.~H. {Chan}},
  \bibinfo{author}{F.~{Courbin}}, \bibinfo{author}{S.~{Hilbert}},
  \bibinfo{author}{O.~{Tihhonova}}, \bibinfo{author}{T.~{Treu}},
  \bibinfo{author}{A.~{Agnello}}, \bibinfo{author}{X.~{Ding}},
  \bibinfo{author}{I.~{Jee}}, \bibinfo{author}{E.~{Komatsu}},
  \bibinfo{author}{A.~J. {Shajib}}, \bibinfo{author}{A.~{Sonnenfeld}},
  \bibinfo{author}{R.~D. {Blandford}}, \bibinfo{author}{L.~V.~E. {Koopmans}},
  \bibinfo{author}{P.~J. {Marshall}}, \bibinfo{author}{G.~{Meylan}},
\newblock \bibinfo{title}{{H0LiCOW - XIII. A 2.4 per cent measurement of
  H$_{0}$ from lensed quasars: 5.3{\ensuremath{\sigma}} tension between early-
  and late-Universe probes}},
\newblock \bibinfo{journal}{\mnras} \bibinfo{volume}{498}
  (\bibinfo{year}{2020}) \bibinfo{pages}{1420--1439}.
  \DOIprefix\doi{10.1093/mnras/stz3094}.
  \href{http://arxiv.org/abs/1907.04869}{{\tt arXiv:1907.04869}}.
%Type = Article
\bibitem[{{Pesce} et~al.(2020){Pesce}, {Braatz}, {Reid}, {Riess}, {Scolnic},
  {Condon}, {Gao}, {Henkel}, {Impellizzeri}, {Kuo}, and
  {Lo}}]{2020ApJ...891L...1P}
\bibinfo{author}{D.~W. {Pesce}}, \bibinfo{author}{J.~A. {Braatz}},
  \bibinfo{author}{M.~J. {Reid}}, \bibinfo{author}{A.~G. {Riess}},
  \bibinfo{author}{D.~{Scolnic}}, \bibinfo{author}{J.~J. {Condon}},
  \bibinfo{author}{F.~{Gao}}, \bibinfo{author}{C.~{Henkel}},
  \bibinfo{author}{C.~M.~V. {Impellizzeri}}, \bibinfo{author}{C.~Y. {Kuo}},
  \bibinfo{author}{K.~Y. {Lo}},
\newblock \bibinfo{title}{{The Megamaser Cosmology Project. XIII. Combined
  Hubble Constant Constraints}},
\newblock \bibinfo{journal}{\apjl} \bibinfo{volume}{891} (\bibinfo{year}{2020})
  \bibinfo{pages}{L1}. \DOIprefix\doi{10.3847/2041-8213/ab75f0}.
  \href{http://arxiv.org/abs/2001.09213}{{\tt arXiv:2001.09213}}.
%Type = Article
\bibitem[{{Freedman} et~al.(2019){Freedman}, {Madore}, {Hatt}, {Hoyt}, {Jang},
  {Beaton}, {Burns}, {Lee}, {Monson}, {Neeley}, {Phillips}, {Rich}, and
  {Seibert}}]{2019ApJ...882...34F}
\bibinfo{author}{W.~L. {Freedman}}, \bibinfo{author}{B.~F. {Madore}},
  \bibinfo{author}{D.~{Hatt}}, \bibinfo{author}{T.~J. {Hoyt}},
  \bibinfo{author}{I.~S. {Jang}}, \bibinfo{author}{R.~L. {Beaton}},
  \bibinfo{author}{C.~R. {Burns}}, \bibinfo{author}{M.~G. {Lee}},
  \bibinfo{author}{A.~J. {Monson}}, \bibinfo{author}{J.~R. {Neeley}},
  \bibinfo{author}{M.~M. {Phillips}}, \bibinfo{author}{J.~A. {Rich}},
  \bibinfo{author}{M.~{Seibert}},
\newblock \bibinfo{title}{{The Carnegie-Chicago Hubble Program. VIII. An
  Independent Determination of the Hubble Constant Based on the Tip of the Red
  Giant Branch}},
\newblock \bibinfo{journal}{\apj} \bibinfo{volume}{882} (\bibinfo{year}{2019})
  \bibinfo{pages}{34}. \DOIprefix\doi{10.3847/1538-4357/ab2f73}.
  \href{http://arxiv.org/abs/1907.05922}{{\tt arXiv:1907.05922}}.
%Type = Article
\bibitem[{{Freedman} et~al.(2024){Freedman}, {Madore}, {Jang}, {Hoyt}, {Lee},
  and {Owens}}]{2024arXiv240806153F}
\bibinfo{author}{W.~L. {Freedman}}, \bibinfo{author}{B.~F. {Madore}},
  \bibinfo{author}{I.~S. {Jang}}, \bibinfo{author}{T.~J. {Hoyt}},
  \bibinfo{author}{A.~J. {Lee}}, \bibinfo{author}{K.~A. {Owens}},
\newblock \bibinfo{title}{{Status Report on the Chicago-Carnegie Hubble Program
  (CCHP): Measurement of the Hubble Constant Using the Hubble and James Webb
  Space Telescopes}},
\newblock \bibinfo{journal}{arXiv e-prints}  (\bibinfo{year}{2024})
  \bibinfo{pages}{arXiv:2408.06153}. \DOIprefix\doi{10.48550/arXiv.2408.06153}.
  \href{http://arxiv.org/abs/2408.06153}{{\tt arXiv:2408.06153}}.
%Type = Article
\bibitem[{{Di Valentino} et~al.(2021){Di Valentino}, {Mena}, {Pan},
  {Visinelli}, {Yang}, {Melchiorri}, {Mota}, {Riess}, and
  {Silk}}]{2021CQGra..38o3001D}
\bibinfo{author}{E.~{Di Valentino}}, \bibinfo{author}{O.~{Mena}},
  \bibinfo{author}{S.~{Pan}}, \bibinfo{author}{L.~{Visinelli}},
  \bibinfo{author}{W.~{Yang}}, \bibinfo{author}{A.~{Melchiorri}},
  \bibinfo{author}{D.~F. {Mota}}, \bibinfo{author}{A.~G. {Riess}},
  \bibinfo{author}{J.~{Silk}},
\newblock \bibinfo{title}{{In the realm of the Hubble tension-a review of
  solutions}},
\newblock \bibinfo{journal}{Classical and Quantum Gravity} \bibinfo{volume}{38}
  (\bibinfo{year}{2021}) \bibinfo{pages}{153001}.
  \DOIprefix\doi{10.1088/1361-6382/ac086d}.
  \href{http://arxiv.org/abs/2103.01183}{{\tt arXiv:2103.01183}}.
%Type = Article
\bibitem[{{Moresco} et~al.(2022){Moresco}, {Amati}, {Amendola}, {Birrer},
  {Blakeslee}, {Cantiello}, {Cimatti}, {Darling}, {Della Valle}, {Fishbach},
  {Grillo}, {Hamaus}, {Holz}, {Izzo}, {Jimenez}, {Lusso}, {Meneghetti},
  {Piedipalumbo}, {Pisani}, {Pourtsidou}, {Pozzetti}, {Quartin}, {Risaliti},
  {Rosati}, and {Verde}}]{2022LRR....25....6M}
\bibinfo{author}{M.~{Moresco}}, \bibinfo{author}{L.~{Amati}},
  \bibinfo{author}{L.~{Amendola}}, \bibinfo{author}{S.~{Birrer}},
  \bibinfo{author}{J.~P. {Blakeslee}}, \bibinfo{author}{M.~{Cantiello}},
  \bibinfo{author}{A.~{Cimatti}}, \bibinfo{author}{J.~{Darling}},
  \bibinfo{author}{M.~{Della Valle}}, \bibinfo{author}{M.~{Fishbach}},
  \bibinfo{author}{C.~{Grillo}}, \bibinfo{author}{N.~{Hamaus}},
  \bibinfo{author}{D.~{Holz}}, \bibinfo{author}{L.~{Izzo}},
  \bibinfo{author}{R.~{Jimenez}}, \bibinfo{author}{E.~{Lusso}},
  \bibinfo{author}{M.~{Meneghetti}}, \bibinfo{author}{E.~{Piedipalumbo}},
  \bibinfo{author}{A.~{Pisani}}, \bibinfo{author}{A.~{Pourtsidou}},
  \bibinfo{author}{L.~{Pozzetti}}, \bibinfo{author}{M.~{Quartin}},
  \bibinfo{author}{G.~{Risaliti}}, \bibinfo{author}{P.~{Rosati}},
  \bibinfo{author}{L.~{Verde}},
\newblock \bibinfo{title}{{Unveiling the Universe with emerging cosmological
  probes}},
\newblock \bibinfo{journal}{\lrr} \bibinfo{volume}{25} (\bibinfo{year}{2022})
  \bibinfo{pages}{6}. \DOIprefix\doi{10.1007/s41114-022-00040-z}.
  \href{http://arxiv.org/abs/2201.07241}{{\tt arXiv:2201.07241}}.
%Type = Article
\bibitem[{Sandage(1962)}]{Sandage1962}
\bibinfo{author}{A.~Sandage},
\newblock \bibinfo{title}{the {Change} of {Redshift} and {Apparent}
  {Luminosity}},
\newblock \bibinfo{journal}{The Astrophysical Journal} \bibinfo{volume}{136}
  (\bibinfo{year}{1962}).
%Type = Article
\bibitem[{{Loeb}(1998)}]{1998ApJ...499L.111L}
\bibinfo{author}{A.~{Loeb}},
\newblock \bibinfo{title}{{Direct Measurement of Cosmological Parameters from
  the Cosmic Deceleration of Extragalactic Objects}},
\newblock \bibinfo{journal}{\apjl} \bibinfo{volume}{499} (\bibinfo{year}{1998})
  \bibinfo{pages}{L111--L114}. \DOIprefix\doi{10.1086/311375}.
  \href{http://arxiv.org/abs/astro-ph/9802122}{{\tt arXiv:astro-ph/9802122}}.
%Type = Book
\bibitem[{Weinberg(2008)}]{Weinberg:2008zzc}
\bibinfo{author}{S.~Weinberg}, \bibinfo{title}{{Cosmology}},
  \bibinfo{publisher}{Oxford Univ. Press}, \bibinfo{address}{Oxford, UK},
  \bibinfo{year}{2008}. \URLprefix
  \url{https://inspirehep.net/literature/794379}.
%Type = Article
\bibitem[{{Liske} et~al.(2008){Liske}, {Grazian}, {Vanzella}, {Dessauges},
  {Viel}, {Pasquini}, {Haehnelt}, {Cristiani}, {Pepe}, {Avila}, {Bonifacio},
  {Bouchy}, {Dekker}, {Delabre}, {D'Odorico}, {D'Odorico}, {Levshakov},
  {Lovis}, {Mayor}, {Molaro}, {Moscardini}, {Murphy}, {Queloz}, {Shaver},
  {Udry}, {Wiklind}, and {Zucker}}]{2008MNRAS.386.1192L}
\bibinfo{author}{J.~{Liske}}, \bibinfo{author}{A.~{Grazian}},
  \bibinfo{author}{E.~{Vanzella}}, \bibinfo{author}{M.~{Dessauges}},
  \bibinfo{author}{M.~{Viel}}, \bibinfo{author}{L.~{Pasquini}},
  \bibinfo{author}{M.~{Haehnelt}}, \bibinfo{author}{S.~{Cristiani}},
  \bibinfo{author}{F.~{Pepe}}, \bibinfo{author}{G.~{Avila}},
  \bibinfo{author}{P.~{Bonifacio}}, \bibinfo{author}{F.~{Bouchy}},
  \bibinfo{author}{H.~{Dekker}}, \bibinfo{author}{B.~{Delabre}},
  \bibinfo{author}{S.~{D'Odorico}}, \bibinfo{author}{V.~{D'Odorico}},
  \bibinfo{author}{S.~{Levshakov}}, \bibinfo{author}{C.~{Lovis}},
  \bibinfo{author}{M.~{Mayor}}, \bibinfo{author}{P.~{Molaro}},
  \bibinfo{author}{L.~{Moscardini}}, \bibinfo{author}{M.~T. {Murphy}},
  \bibinfo{author}{D.~{Queloz}}, \bibinfo{author}{P.~{Shaver}},
  \bibinfo{author}{S.~{Udry}}, \bibinfo{author}{T.~{Wiklind}},
  \bibinfo{author}{S.~{Zucker}},
\newblock \bibinfo{title}{{Cosmic dynamics in the era of Extremely Large
  Telescopes}},
\newblock \bibinfo{journal}{\mnras} \bibinfo{volume}{386}
  (\bibinfo{year}{2008}) \bibinfo{pages}{1192--1218}.
  \DOIprefix\doi{10.1111/j.1365-2966.2008.13090.x}.
  \href{http://arxiv.org/abs/0802.1532}{{\tt arXiv:0802.1532}}.
%Type = Article
\bibitem[{{Martins} et~al.(2024){Martins}, {Cooke}, {Liske}, {Murphy},
  {Noterdaeme}, {Schmidt}, {Alcaniz}, {Alves}, {Balashev}, {Cristiani}, {Di
  Marcantonio}, {G{\'e}nova Santos}, {Gon{\c{c}}alves}, {Gonz{\'a}lez
  Hern{\'a}ndez}, {Maiolino}, {Marconi}, {Marques}, {Melo e Sousa}, {Nunes},
  {Origlia}, {P{\'e}roux}, {Vinzl}, and {Zanutta}}]{2024ExA....57....5M}
\bibinfo{author}{C.~J.~A.~P. {Martins}}, \bibinfo{author}{R.~{Cooke}},
  \bibinfo{author}{J.~{Liske}}, \bibinfo{author}{M.~T. {Murphy}},
  \bibinfo{author}{P.~{Noterdaeme}}, \bibinfo{author}{T.~M. {Schmidt}},
  \bibinfo{author}{J.~S. {Alcaniz}}, \bibinfo{author}{C.~S. {Alves}},
  \bibinfo{author}{S.~{Balashev}}, \bibinfo{author}{S.~{Cristiani}},
  \bibinfo{author}{P.~{Di Marcantonio}}, \bibinfo{author}{R.~{G{\'e}nova
  Santos}}, \bibinfo{author}{R.~S. {Gon{\c{c}}alves}}, \bibinfo{author}{J.~I.
  {Gonz{\'a}lez Hern{\'a}ndez}}, \bibinfo{author}{R.~{Maiolino}},
  \bibinfo{author}{A.~{Marconi}}, \bibinfo{author}{C.~M.~J. {Marques}},
  \bibinfo{author}{M.~A.~F. {Melo e Sousa}}, \bibinfo{author}{N.~J. {Nunes}},
  \bibinfo{author}{L.~{Origlia}}, \bibinfo{author}{C.~{P{\'e}roux}},
  \bibinfo{author}{S.~{Vinzl}}, \bibinfo{author}{A.~{Zanutta}},
\newblock \bibinfo{title}{{Cosmology and fundamental physics with the ELT-ANDES
  spectrograph}},
\newblock \bibinfo{journal}{Experimental Astronomy} \bibinfo{volume}{57}
  (\bibinfo{year}{2024}) \bibinfo{pages}{5}.
  \DOIprefix\doi{10.1007/s10686-024-09928-w}.
  \href{http://arxiv.org/abs/2311.16274}{{\tt arXiv:2311.16274}}.
%Type = Article
\bibitem[{{Darling}(2012)}]{2012ApJ...761L..26D}
\bibinfo{author}{J.~{Darling}},
\newblock \bibinfo{title}{{Toward a Direct Measurement of the Cosmic
  Acceleration}},
\newblock \bibinfo{journal}{\apjl} \bibinfo{volume}{761} (\bibinfo{year}{2012})
  \bibinfo{pages}{L26}. \DOIprefix\doi{10.1088/2041-8205/761/2/L26}.
  \href{http://arxiv.org/abs/1211.4585}{{\tt arXiv:1211.4585}}.
%Type = Article
\bibitem[{{Yu} et~al.(2014){Yu}, {Zhang}, and {Pen}}]{2014PhRvL.113d1303Y}
\bibinfo{author}{H.-R. {Yu}}, \bibinfo{author}{T.-J. {Zhang}},
  \bibinfo{author}{U.-L. {Pen}},
\newblock \bibinfo{title}{{Method for Direct Measurement of Cosmic Acceleration
  by 21-cm Absorption Systems}},
\newblock \bibinfo{journal}{\prl} \bibinfo{volume}{113} (\bibinfo{year}{2014})
  \bibinfo{pages}{041303}. \DOIprefix\doi{10.1103/PhysRevLett.113.041303}.
  \href{http://arxiv.org/abs/1311.2363}{{\tt arXiv:1311.2363}}.
%Type = Inproceedings
\bibitem[{{Kloeckner} et~al.(2015){Kloeckner}, {Obreschkow}, {Martins},
  {Raccanelli}, {Champion}, {Roy}, {Lobanov}, {Wagner}, and
  {Keller}}]{2015aska.confE..27K}
\bibinfo{author}{H.~R. {Kloeckner}}, \bibinfo{author}{D.~{Obreschkow}},
  \bibinfo{author}{C.~{Martins}}, \bibinfo{author}{A.~{Raccanelli}},
  \bibinfo{author}{D.~{Champion}}, \bibinfo{author}{A.~L. {Roy}},
  \bibinfo{author}{A.~{Lobanov}}, \bibinfo{author}{J.~{Wagner}},
  \bibinfo{author}{R.~{Keller}},
\newblock \bibinfo{title}{{Real time cosmology - A direct measure of the
  expansion rate of the Universe with the SKA}},
\newblock in: \bibinfo{booktitle}{Advancing Astrophysics with the Square
  Kilometre Array (AASKA14)}, \bibinfo{year}{2015}, p.~\bibinfo{pages}{27}.
  \DOIprefix\doi{10.22323/1.215.0027}.
  \href{http://arxiv.org/abs/1501.03822}{{\tt arXiv:1501.03822}}.
%Type = Article
\bibitem[{{Jiao} et~al.(2020){Jiao}, {Zhang}, {Zhang}, {Yu}, {Zhu}, and
  {Li}}]{2020JCAP...01..054J}
\bibinfo{author}{K.~{Jiao}}, \bibinfo{author}{J.-C. {Zhang}},
  \bibinfo{author}{T.-J. {Zhang}}, \bibinfo{author}{H.-R. {Yu}},
  \bibinfo{author}{M.~{Zhu}}, \bibinfo{author}{D.~{Li}},
\newblock \bibinfo{title}{{Toward a direct measurement of the cosmic
  acceleration: roadmap and forecast on FAST}},
\newblock \bibinfo{journal}{\jcap} \bibinfo{volume}{2020}
  (\bibinfo{year}{2020}) \bibinfo{pages}{054}.
  \DOIprefix\doi{10.1088/1475-7516/2020/01/054}.
  \href{http://arxiv.org/abs/1905.01184}{{\tt arXiv:1905.01184}}.
%Type = Article
\bibitem[{{Cooke}(2020)}]{2020MNRAS.492.2044C}
\bibinfo{author}{R.~{Cooke}},
\newblock \bibinfo{title}{{The ACCELERATION programme: I. Cosmology with the
  redshift drift}},
\newblock \bibinfo{journal}{\mnras} \bibinfo{volume}{492}
  (\bibinfo{year}{2020}) \bibinfo{pages}{2044--2057}.
  \DOIprefix\doi{10.1093/mnras/stz3465}.
  \href{http://arxiv.org/abs/1912.04983}{{\tt arXiv:1912.04983}}.
%Type = Article
\bibitem[{{Capozziello} et~al.(2019){Capozziello}, {D'Agostino}, and
  {Luongo}}]{2019GReGr..51....2C}
\bibinfo{author}{S.~{Capozziello}}, \bibinfo{author}{R.~{D'Agostino}},
  \bibinfo{author}{O.~{Luongo}},
\newblock \bibinfo{title}{{Kinematic model-independent reconstruction of
  Palatini f( R) cosmology}},
\newblock \bibinfo{journal}{\grg} \bibinfo{volume}{51} (\bibinfo{year}{2019})
  \bibinfo{pages}{2}. \DOIprefix\doi{10.1007/s10714-018-2483-0}.
  \href{http://arxiv.org/abs/1806.06385}{{\tt arXiv:1806.06385}}.
%Type = Article
\bibitem[{{Shafieloo} et~al.(2012){Shafieloo}, {Kim}, and
  {Linder}}]{2012PhRvD..85l3530S}
\bibinfo{author}{A.~{Shafieloo}}, \bibinfo{author}{A.~G. {Kim}},
  \bibinfo{author}{E.~V. {Linder}},
\newblock \bibinfo{title}{{Gaussian process cosmography}},
\newblock \bibinfo{journal}{\prd} \bibinfo{volume}{85} (\bibinfo{year}{2012})
  \bibinfo{pages}{123530}. \DOIprefix\doi{10.1103/PhysRevD.85.123530}.
  \href{http://arxiv.org/abs/1204.2272}{{\tt arXiv:1204.2272}}.
%Type = Article
\bibitem[{{Zhang} et~al.(2024){Zhang}, {Hu}, {Jiao}, {Wang}, {Xie}, {Yu},
  {Zhao}, and {Zhang}}]{2024ApJS..270...23Z}
\bibinfo{author}{J.-C. {Zhang}}, \bibinfo{author}{Y.~{Hu}},
  \bibinfo{author}{K.~{Jiao}}, \bibinfo{author}{H.-F. {Wang}},
  \bibinfo{author}{Y.-B. {Xie}}, \bibinfo{author}{B.~{Yu}},
  \bibinfo{author}{L.-L. {Zhao}}, \bibinfo{author}{T.-J. {Zhang}},
\newblock \bibinfo{title}{{A Nonparametric Reconstruction of the Hubble
  Parameter H(z) Based on Radial Basis Function Neural Networks}},
\newblock \bibinfo{journal}{\apjs} \bibinfo{volume}{270} (\bibinfo{year}{2024})
  \bibinfo{pages}{23}. \DOIprefix\doi{10.3847/1538-4365/ad0f1e}.
  \href{http://arxiv.org/abs/2311.13938}{{\tt arXiv:2311.13938}}.
%Type = Article
\bibitem[{{Linder}(2017)}]{2017APh....86...41L}
\bibinfo{author}{E.~V. {Linder}},
\newblock \bibinfo{title}{{Cosmic growth and expansion conjoined}},
\newblock \bibinfo{journal}{\ap} \bibinfo{volume}{86} (\bibinfo{year}{2017})
  \bibinfo{pages}{41--45}. \DOIprefix\doi{10.1016/j.astropartphys.2016.11.002}.
  \href{http://arxiv.org/abs/1610.05321}{{\tt arXiv:1610.05321}}.
%Type = Article
\bibitem[{{Moresco} and {Marulli}(2017)}]{2017MNRAS.471L..82M}
\bibinfo{author}{M.~{Moresco}}, \bibinfo{author}{F.~{Marulli}},
\newblock \bibinfo{title}{{Cosmological constraints from a joint analysis of
  cosmic growth and expansion}},
\newblock \bibinfo{journal}{\mnras} \bibinfo{volume}{471}
  (\bibinfo{year}{2017}) \bibinfo{pages}{L82--L86}.
  \DOIprefix\doi{10.1093/mnrasl/slx112}.
  \href{http://arxiv.org/abs/1705.07903}{{\tt arXiv:1705.07903}}.
%Type = Article
\bibitem[{{Moresco} et~al.(2020){Moresco}, {Jimenez}, {Verde}, {Cimatti}, and
  {Pozzetti}}]{2020ApJ...898...82M}
\bibinfo{author}{M.~{Moresco}}, \bibinfo{author}{R.~{Jimenez}},
  \bibinfo{author}{L.~{Verde}}, \bibinfo{author}{A.~{Cimatti}},
  \bibinfo{author}{L.~{Pozzetti}},
\newblock \bibinfo{title}{{Setting the Stage for Cosmic Chronometers. II.
  Impact of Stellar Population Synthesis Models Systematics and Full Covariance
  Matrix}},
\newblock \bibinfo{journal}{\apj} \bibinfo{volume}{898} (\bibinfo{year}{2020})
  \bibinfo{pages}{82}. \DOIprefix\doi{10.3847/1538-4357/ab9eb0}.
  \href{http://arxiv.org/abs/2003.07362}{{\tt arXiv:2003.07362}}.
%Type = Misc
\bibitem[{{Foreman-Mackey} et~al.(2013){Foreman-Mackey}, {Conley}, {Meierjurgen
  Farr}, {Hogg}, {Lang}, {Marshall}, {Price-Whelan}, {Sanders}, and
  {Zuntz}}]{2013ascl.soft03002F}
\bibinfo{author}{D.~{Foreman-Mackey}}, \bibinfo{author}{A.~{Conley}},
  \bibinfo{author}{W.~{Meierjurgen Farr}}, \bibinfo{author}{D.~W. {Hogg}},
  \bibinfo{author}{D.~{Lang}}, \bibinfo{author}{P.~{Marshall}},
  \bibinfo{author}{A.~{Price-Whelan}}, \bibinfo{author}{J.~{Sanders}},
  \bibinfo{author}{J.~{Zuntz}}, \bibinfo{title}{{emcee: The MCMC Hammer}},
  \bibinfo{howpublished}{Astrophysics Source Code Library, record
  ascl:1303.002}, \bibinfo{year}{2013}.
%Type = Article
\bibitem[{{Zhang} et~al.(2014){Zhang}, {Zhang}, {Yuan}, {Liu}, {Zhang}, and
  {Sun}}]{2014RAA....14.1221Z}
\bibinfo{author}{C.~{Zhang}}, \bibinfo{author}{H.~{Zhang}},
  \bibinfo{author}{S.~{Yuan}}, \bibinfo{author}{S.~{Liu}},
  \bibinfo{author}{T.-J. {Zhang}}, \bibinfo{author}{Y.-C. {Sun}},
\newblock \bibinfo{title}{{Four new observational H(z) data from luminous red
  galaxies in the Sloan Digital Sky Survey data release seven}},
\newblock \bibinfo{journal}{Research in Astronomy and Astrophysics}
  \bibinfo{volume}{14} (\bibinfo{year}{2014}) \bibinfo{pages}{1221--1233}.
  \DOIprefix\doi{10.1088/1674-4527/14/10/002}.
  \href{http://arxiv.org/abs/1207.4541}{{\tt arXiv:1207.4541}}.
%Type = Article
\bibitem[{{Jimenez} et~al.(2003){Jimenez}, {Verde}, {Treu}, and
  {Stern}}]{2003ApJ...593..622J}
\bibinfo{author}{R.~{Jimenez}}, \bibinfo{author}{L.~{Verde}},
  \bibinfo{author}{T.~{Treu}}, \bibinfo{author}{D.~{Stern}},
\newblock \bibinfo{title}{{Constraints on the Equation of State of Dark Energy
  and the Hubble Constant from Stellar Ages and the Cosmic Microwave
  Background}},
\newblock \bibinfo{journal}{\apj} \bibinfo{volume}{593} (\bibinfo{year}{2003})
  \bibinfo{pages}{622--629}. \DOIprefix\doi{10.1086/376595}.
  \href{http://arxiv.org/abs/astro-ph/0302560}{{\tt arXiv:astro-ph/0302560}}.
%Type = Article
\bibitem[{{Simon} et~al.(2005){Simon}, {Verde}, and
  {Jimenez}}]{2005PhRvD..71l3001S}
\bibinfo{author}{J.~{Simon}}, \bibinfo{author}{L.~{Verde}},
  \bibinfo{author}{R.~{Jimenez}},
\newblock \bibinfo{title}{{Constraints on the redshift dependence of the dark
  energy potential}},
\newblock \bibinfo{journal}{\prd} \bibinfo{volume}{71} (\bibinfo{year}{2005})
  \bibinfo{pages}{123001}. \DOIprefix\doi{10.1103/PhysRevD.71.123001}.
  \href{http://arxiv.org/abs/astro-ph/0412269}{{\tt arXiv:astro-ph/0412269}}.
%Type = Article
\bibitem[{{Moresco} et~al.(2012){Moresco}, {Cimatti}, {Jimenez}, {Pozzetti},
  {Zamorani}, {Bolzonella}, {Dunlop}, {Lamareille}, {Mignoli}, {Pearce},
  {Rosati}, {Stern}, {Verde}, {Zucca}, {Carollo}, {Contini}, {Kneib}, {Le
  F{\`e}vre}, {Lilly}, {Mainieri}, {Renzini}, {Scodeggio}, {Balestra}, {Gobat},
  {McLure}, {Bardelli}, {Bongiorno}, {Caputi}, {Cucciati}, {de la Torre}, {de
  Ravel}, {Franzetti}, {Garilli}, {Iovino}, {Kampczyk}, {Knobel},
  {Kova{\v{c}}}, {Le Borgne}, {Le Brun}, {Maier}, {Pell{\'o}}, {Peng},
  {Perez-Montero}, {Presotto}, {Silverman}, {Tanaka}, {Tasca}, {Tresse},
  {Vergani}, {Almaini}, {Barnes}, {Bordoloi}, {Bradshaw}, {Cappi}, {Chuter},
  {Cirasuolo}, {Coppa}, {Diener}, {Foucaud}, {Hartley}, {Kamionkowski},
  {Koekemoer}, {L{\'o}pez-Sanjuan}, {McCracken}, {Nair}, {Oesch}, {Stanford},
  and {Welikala}}]{2012JCAP...08..006M}
\bibinfo{author}{M.~{Moresco}}, \bibinfo{author}{A.~{Cimatti}},
  \bibinfo{author}{R.~{Jimenez}}, \bibinfo{author}{L.~{Pozzetti}},
  \bibinfo{author}{G.~{Zamorani}}, \bibinfo{author}{M.~{Bolzonella}},
  \bibinfo{author}{J.~{Dunlop}}, \bibinfo{author}{F.~{Lamareille}},
  \bibinfo{author}{M.~{Mignoli}}, \bibinfo{author}{H.~{Pearce}},
  \bibinfo{author}{P.~{Rosati}}, \bibinfo{author}{D.~{Stern}},
  \bibinfo{author}{L.~{Verde}}, \bibinfo{author}{E.~{Zucca}},
  \bibinfo{author}{C.~M. {Carollo}}, \bibinfo{author}{T.~{Contini}},
  \bibinfo{author}{J.~P. {Kneib}}, \bibinfo{author}{O.~{Le F{\`e}vre}},
  \bibinfo{author}{S.~J. {Lilly}}, \bibinfo{author}{V.~{Mainieri}},
  \bibinfo{author}{A.~{Renzini}}, \bibinfo{author}{M.~{Scodeggio}},
  \bibinfo{author}{I.~{Balestra}}, \bibinfo{author}{R.~{Gobat}},
  \bibinfo{author}{R.~{McLure}}, \bibinfo{author}{S.~{Bardelli}},
  \bibinfo{author}{A.~{Bongiorno}}, \bibinfo{author}{K.~{Caputi}},
  \bibinfo{author}{O.~{Cucciati}}, \bibinfo{author}{S.~{de la Torre}},
  \bibinfo{author}{L.~{de Ravel}}, \bibinfo{author}{P.~{Franzetti}},
  \bibinfo{author}{B.~{Garilli}}, \bibinfo{author}{A.~{Iovino}},
  \bibinfo{author}{P.~{Kampczyk}}, \bibinfo{author}{C.~{Knobel}},
  \bibinfo{author}{K.~{Kova{\v{c}}}}, \bibinfo{author}{J.~F. {Le Borgne}},
  \bibinfo{author}{V.~{Le Brun}}, \bibinfo{author}{C.~{Maier}},
  \bibinfo{author}{R.~{Pell{\'o}}}, \bibinfo{author}{Y.~{Peng}},
  \bibinfo{author}{E.~{Perez-Montero}}, \bibinfo{author}{V.~{Presotto}},
  \bibinfo{author}{J.~D. {Silverman}}, \bibinfo{author}{M.~{Tanaka}},
  \bibinfo{author}{L.~A.~M. {Tasca}}, \bibinfo{author}{L.~{Tresse}},
  \bibinfo{author}{D.~{Vergani}}, \bibinfo{author}{O.~{Almaini}},
  \bibinfo{author}{L.~{Barnes}}, \bibinfo{author}{R.~{Bordoloi}},
  \bibinfo{author}{E.~{Bradshaw}}, \bibinfo{author}{A.~{Cappi}},
  \bibinfo{author}{R.~{Chuter}}, \bibinfo{author}{M.~{Cirasuolo}},
  \bibinfo{author}{G.~{Coppa}}, \bibinfo{author}{C.~{Diener}},
  \bibinfo{author}{S.~{Foucaud}}, \bibinfo{author}{W.~{Hartley}},
  \bibinfo{author}{M.~{Kamionkowski}}, \bibinfo{author}{A.~M. {Koekemoer}},
  \bibinfo{author}{C.~{L{\'o}pez-Sanjuan}}, \bibinfo{author}{H.~J.
  {McCracken}}, \bibinfo{author}{P.~{Nair}}, \bibinfo{author}{P.~{Oesch}},
  \bibinfo{author}{A.~{Stanford}}, \bibinfo{author}{N.~{Welikala}},
\newblock \bibinfo{title}{{Improved constraints on the expansion rate of the
  Universe up to z \raisebox{-0.5ex}\textasciitilde 1.1 from the spectroscopic
  evolution of cosmic chronometers}},
\newblock \bibinfo{journal}{\jcap} \bibinfo{volume}{2012}
  (\bibinfo{year}{2012}) \bibinfo{pages}{006}.
  \DOIprefix\doi{10.1088/1475-7516/2012/08/006}.
  \href{http://arxiv.org/abs/1201.3609}{{\tt arXiv:1201.3609}}.
%Type = Article
\bibitem[{{Moresco} et~al.(2016){Moresco}, {Pozzetti}, {Cimatti}, {Jimenez},
  {Maraston}, {Verde}, {Thomas}, {Citro}, {Tojeiro}, and
  {Wilkinson}}]{2016JCAP...05..014M}
\bibinfo{author}{M.~{Moresco}}, \bibinfo{author}{L.~{Pozzetti}},
  \bibinfo{author}{A.~{Cimatti}}, \bibinfo{author}{R.~{Jimenez}},
  \bibinfo{author}{C.~{Maraston}}, \bibinfo{author}{L.~{Verde}},
  \bibinfo{author}{D.~{Thomas}}, \bibinfo{author}{A.~{Citro}},
  \bibinfo{author}{R.~{Tojeiro}}, \bibinfo{author}{D.~{Wilkinson}},
\newblock \bibinfo{title}{{A 6\% measurement of the Hubble parameter at
  z\raisebox{-0.5ex}\textasciitilde0.45: direct evidence of the epoch of cosmic
  re-acceleration}},
\newblock \bibinfo{journal}{\jcap} \bibinfo{volume}{2016}
  (\bibinfo{year}{2016}) \bibinfo{pages}{014}.
  \DOIprefix\doi{10.1088/1475-7516/2016/05/014}.
  \href{http://arxiv.org/abs/1601.01701}{{\tt arXiv:1601.01701}}.
%Type = Article
\bibitem[{{Ratsimbazafy} et~al.(2017){Ratsimbazafy}, {Loubser}, {Crawford},
  {Cress}, {Bassett}, {Nichol}, and {V{\"a}is{\"a}nen}}]{2017MNRAS.467.3239R}
\bibinfo{author}{A.~L. {Ratsimbazafy}}, \bibinfo{author}{S.~I. {Loubser}},
  \bibinfo{author}{S.~M. {Crawford}}, \bibinfo{author}{C.~M. {Cress}},
  \bibinfo{author}{B.~A. {Bassett}}, \bibinfo{author}{R.~C. {Nichol}},
  \bibinfo{author}{P.~{V{\"a}is{\"a}nen}},
\newblock \bibinfo{title}{{Age-dating luminous red galaxies observed with the
  Southern African Large Telescope}},
\newblock \bibinfo{journal}{\mnras} \bibinfo{volume}{467}
  (\bibinfo{year}{2017}) \bibinfo{pages}{3239--3254}.
  \DOIprefix\doi{10.1093/mnras/stx301}.
  \href{http://arxiv.org/abs/1702.00418}{{\tt arXiv:1702.00418}}.
%Type = Article
\bibitem[{{Stern} et~al.(2010){Stern}, {Jimenez}, {Verde}, {Kamionkowski}, and
  {Stanford}}]{2010JCAP...02..008S}
\bibinfo{author}{D.~{Stern}}, \bibinfo{author}{R.~{Jimenez}},
  \bibinfo{author}{L.~{Verde}}, \bibinfo{author}{M.~{Kamionkowski}},
  \bibinfo{author}{S.~A. {Stanford}},
\newblock \bibinfo{title}{{Cosmic chronometers: constraining the equation of
  state of dark energy. I: H(z) measurements}},
\newblock \bibinfo{journal}{\jcap} \bibinfo{volume}{2010}
  (\bibinfo{year}{2010}) \bibinfo{pages}{008}.
  \DOIprefix\doi{10.1088/1475-7516/2010/02/008}.
  \href{http://arxiv.org/abs/0907.3149}{{\tt arXiv:0907.3149}}.
%Type = Article
\bibitem[{{Loubser} et~al.(2025){Loubser}, {Alabi}, {Hilton}, {Ma}, {Tang},
  {Hatamkhani}, {Cress}, {Skelton}, and {Nkosi}}]{2025MNRAS.540.3135L}
\bibinfo{author}{S.~I. {Loubser}}, \bibinfo{author}{A.~B. {Alabi}},
  \bibinfo{author}{M.~{Hilton}}, \bibinfo{author}{Y.-Z. {Ma}},
  \bibinfo{author}{X.~{Tang}}, \bibinfo{author}{N.~{Hatamkhani}},
  \bibinfo{author}{C.~{Cress}}, \bibinfo{author}{R.~E. {Skelton}},
  \bibinfo{author}{S.~A. {Nkosi}},
\newblock \bibinfo{title}{{An independent estimate of H(z) at z = 0.5 from the
  stellar ages of brightest cluster galaxies}},
\newblock \bibinfo{journal}{\mnras} \bibinfo{volume}{540}
  (\bibinfo{year}{2025}) \bibinfo{pages}{3135--3149}.
  \DOIprefix\doi{10.1093/mnras/staf915}.
  \href{http://arxiv.org/abs/2506.03836}{{\tt arXiv:2506.03836}}.
%Type = Article
\bibitem[{{Borghi} et~al.(2022){Borghi}, {Moresco}, and
  {Cimatti}}]{2022ApJ...928L...4B}
\bibinfo{author}{N.~{Borghi}}, \bibinfo{author}{M.~{Moresco}},
  \bibinfo{author}{A.~{Cimatti}},
\newblock \bibinfo{title}{{Toward a Better Understanding of Cosmic
  Chronometers: A New Measurement of H(z) at z 0.7}},
\newblock \bibinfo{journal}{\apjl} \bibinfo{volume}{928} (\bibinfo{year}{2022})
  \bibinfo{pages}{L4}. \DOIprefix\doi{10.3847/2041-8213/ac3fb2}.
  \href{http://arxiv.org/abs/2110.04304}{{\tt arXiv:2110.04304}}.
%Type = Article
\bibitem[{{Jimenez} et~al.(2023){Jimenez}, {Moresco}, {Verde}, and
  {Wandelt}}]{2023JCAP...11..047J}
\bibinfo{author}{R.~{Jimenez}}, \bibinfo{author}{M.~{Moresco}},
  \bibinfo{author}{L.~{Verde}}, \bibinfo{author}{B.~D. {Wandelt}},
\newblock \bibinfo{title}{{Cosmic chronometers with photometry: a new path to
  H(z)}},
\newblock \bibinfo{journal}{\jcap} \bibinfo{volume}{2023}
  (\bibinfo{year}{2023}) \bibinfo{pages}{047}.
  \DOIprefix\doi{10.1088/1475-7516/2023/11/047}.
  \href{http://arxiv.org/abs/2306.11425}{{\tt arXiv:2306.11425}}.
%Type = Article
\bibitem[{{Jiao} et~al.(2023){Jiao}, {Borghi}, {Moresco}, and
  {Zhang}}]{2023ApJS..265...48J}
\bibinfo{author}{K.~{Jiao}}, \bibinfo{author}{N.~{Borghi}},
  \bibinfo{author}{M.~{Moresco}}, \bibinfo{author}{T.-J. {Zhang}},
\newblock \bibinfo{title}{{New Observational H(z) Data from Full-spectrum
  Fitting of Cosmic Chronometers in the LEGA-C Survey}},
\newblock \bibinfo{journal}{\apjs} \bibinfo{volume}{265} (\bibinfo{year}{2023})
  \bibinfo{pages}{48}. \DOIprefix\doi{10.3847/1538-4365/acbc77}.
  \href{http://arxiv.org/abs/2205.05701}{{\tt arXiv:2205.05701}}.
%Type = Article
\bibitem[{{Tomasetti} et~al.(2023){Tomasetti}, {Moresco}, {Borghi}, {Jiao},
  {Cimatti}, {Pozzetti}, {Carnall}, {McLure}, and
  {Pentericci}}]{2023AA...679A..96T}
\bibinfo{author}{E.~{Tomasetti}}, \bibinfo{author}{M.~{Moresco}},
  \bibinfo{author}{N.~{Borghi}}, \bibinfo{author}{K.~{Jiao}},
  \bibinfo{author}{A.~{Cimatti}}, \bibinfo{author}{L.~{Pozzetti}},
  \bibinfo{author}{A.~C. {Carnall}}, \bibinfo{author}{R.~J. {McLure}},
  \bibinfo{author}{L.~{Pentericci}},
\newblock \bibinfo{title}{{A new measurement of the expansion history of the
  Universe at z = 1.26 with cosmic chronometers in VANDELS}},
\newblock \bibinfo{journal}{\aap} \bibinfo{volume}{679} (\bibinfo{year}{2023})
  \bibinfo{pages}{A96}. \DOIprefix\doi{10.1051/0004-6361/202346992}.
  \href{http://arxiv.org/abs/2305.16387}{{\tt arXiv:2305.16387}}.
%Type = Article
\bibitem[{{Moresco}(2015)}]{2015MNRAS.450L..16M}
\bibinfo{author}{M.~{Moresco}},
\newblock \bibinfo{title}{{Raising the bar: new constraints on the Hubble
  parameter with cosmic chronometers at z \raisebox{-0.5ex}\textasciitilde
  2.}},
\newblock \bibinfo{journal}{\mnras} \bibinfo{volume}{450}
  (\bibinfo{year}{2015}) \bibinfo{pages}{L16--L20}.
  \DOIprefix\doi{10.1093/mnrasl/slv037}.
  \href{http://arxiv.org/abs/1503.01116}{{\tt arXiv:1503.01116}}.
%Type = Article
\bibitem[{{Ahlstr{\"o}m Kjerrgren} and
  {M{\"o}rtsell}(2023)}]{2023MNRAS.518..585A}
\bibinfo{author}{A.~{Ahlstr{\"o}m Kjerrgren}},
  \bibinfo{author}{E.~{M{\"o}rtsell}},
\newblock \bibinfo{title}{{On the use of galaxies as clocks and the universal
  expansion}},
\newblock \bibinfo{journal}{\mnras} \bibinfo{volume}{518}
  (\bibinfo{year}{2023}) \bibinfo{pages}{585--591}.
  \DOIprefix\doi{10.1093/mnras/stac1978}.
  \href{http://arxiv.org/abs/2106.11317}{{\tt arXiv:2106.11317}}.
%Type = Article
\bibitem[{{Moresco}(2023)}]{2023arXiv230709501M}
\bibinfo{author}{M.~{Moresco}},
\newblock \bibinfo{title}{{Addressing the Hubble tension with cosmic
  chronometers}},
\newblock \bibinfo{journal}{arXiv e-prints}  (\bibinfo{year}{2023})
  \bibinfo{pages}{arXiv:2307.09501}. \DOIprefix\doi{10.48550/arXiv.2307.09501}.
  \href{http://arxiv.org/abs/2307.09501}{{\tt arXiv:2307.09501}}.
%Type = Article
\bibitem[{{Moresco} et~al.(2018){Moresco}, {Jimenez}, {Verde}, {Pozzetti},
  {Cimatti}, and {Citro}}]{2018ApJ...868...84M}
\bibinfo{author}{M.~{Moresco}}, \bibinfo{author}{R.~{Jimenez}},
  \bibinfo{author}{L.~{Verde}}, \bibinfo{author}{L.~{Pozzetti}},
  \bibinfo{author}{A.~{Cimatti}}, \bibinfo{author}{A.~{Citro}},
\newblock \bibinfo{title}{{Setting the Stage for Cosmic Chronometers. I.
  Assessing the Impact of Young Stellar Populations on Hubble Parameter
  Measurements}},
\newblock \bibinfo{journal}{\apj} \bibinfo{volume}{868} (\bibinfo{year}{2018})
  \bibinfo{pages}{84}. \DOIprefix\doi{10.3847/1538-4357/aae829}.
  \href{http://arxiv.org/abs/1804.05864}{{\tt arXiv:1804.05864}}.
%Type = Article
\bibitem[{{Lu} et~al.(2022){Lu}, {Jiao}, {Zhang}, {Zhang}, and
  {Zhu}}]{2022PDU....3701088L}
\bibinfo{author}{C.-Z. {Lu}}, \bibinfo{author}{K.~{Jiao}},
  \bibinfo{author}{T.~{Zhang}}, \bibinfo{author}{T.-J. {Zhang}},
  \bibinfo{author}{M.~{Zhu}},
\newblock \bibinfo{title}{{Toward a direct measurement of the cosmic
  acceleration: The first preparation with FAST}},
\newblock \bibinfo{journal}{Physics of the Dark Universe} \bibinfo{volume}{37}
  (\bibinfo{year}{2022}) \bibinfo{pages}{101088}.
  \DOIprefix\doi{10.1016/j.dark.2022.101088}.
  \href{http://arxiv.org/abs/2111.10240}{{\tt arXiv:2111.10240}}.
%Type = Article
\bibitem[{{Kang} et~al.(2024){Kang}, {Lu}, {Zhang}, and
  {Zhu}}]{2024RAA....24g5002K}
\bibinfo{author}{J.~{Kang}}, \bibinfo{author}{C.-Z. {Lu}},
  \bibinfo{author}{T.-J. {Zhang}}, \bibinfo{author}{M.~{Zhu}},
\newblock \bibinfo{title}{{Toward a Direct Measurement of the Cosmic
  Acceleration: The Pilot Observation of HI 21 cm Absorption Line at FAST}},
\newblock \bibinfo{journal}{Research in Astronomy and Astrophysics}
  \bibinfo{volume}{24} (\bibinfo{year}{2024}) \bibinfo{pages}{075002}.
  \DOIprefix\doi{10.1088/1674-4527/ad48d1}.
  \href{http://arxiv.org/abs/2308.08851}{{\tt arXiv:2308.08851}}.
%Type = Article
\bibitem[{{Cristiani} et~al.(2023){Cristiani}, {Porru}, {Guarneri},
  {Calderone}, {Boutsia}, {Grazian}, {Cupani}, {D'Odorico}, {Fontanot},
  {Martins}, {Marques}, {Maitra}, and {Trost}}]{2023MNRAS.522.2019C}
\bibinfo{author}{S.~{Cristiani}}, \bibinfo{author}{M.~{Porru}},
  \bibinfo{author}{F.~{Guarneri}}, \bibinfo{author}{G.~{Calderone}},
  \bibinfo{author}{K.~{Boutsia}}, \bibinfo{author}{A.~{Grazian}},
  \bibinfo{author}{G.~{Cupani}}, \bibinfo{author}{V.~{D'Odorico}},
  \bibinfo{author}{F.~{Fontanot}}, \bibinfo{author}{C.~J.~A.~P. {Martins}},
  \bibinfo{author}{C.~M.~J. {Marques}}, \bibinfo{author}{S.~{Maitra}},
  \bibinfo{author}{A.~{Trost}},
\newblock \bibinfo{title}{{Spectroscopy of QUBRICS quasar candidates: 1672 new
  redshifts and a golden sample for the Sandage test of the redshift drift}},
\newblock \bibinfo{journal}{\mnras} \bibinfo{volume}{522}
  (\bibinfo{year}{2023}) \bibinfo{pages}{2019--2028}.
  \DOIprefix\doi{10.1093/mnras/stad1007}.
  \href{http://arxiv.org/abs/2304.00362}{{\tt arXiv:2304.00362}}.
%Type = Article
\bibitem[{{Boutsia} et~al.(2020){Boutsia}, {Grazian}, {Calderone}, {Cristiani},
  {Cupani}, {Guarneri}, {Fontanot}, {Amorin}, {D'Odorico}, {Giallongo},
  {Salvato}, {Omizzolo}, {Romano}, and {Menci}}]{2020ApJS..250...26B}
\bibinfo{author}{K.~{Boutsia}}, \bibinfo{author}{A.~{Grazian}},
  \bibinfo{author}{G.~{Calderone}}, \bibinfo{author}{S.~{Cristiani}},
  \bibinfo{author}{G.~{Cupani}}, \bibinfo{author}{F.~{Guarneri}},
  \bibinfo{author}{F.~{Fontanot}}, \bibinfo{author}{R.~{Amorin}},
  \bibinfo{author}{V.~{D'Odorico}}, \bibinfo{author}{E.~{Giallongo}},
  \bibinfo{author}{M.~{Salvato}}, \bibinfo{author}{A.~{Omizzolo}},
  \bibinfo{author}{M.~{Romano}}, \bibinfo{author}{N.~{Menci}},
\newblock \bibinfo{title}{{The Spectroscopic Follow-up of the QUBRICS Bright
  Quasar Survey}},
\newblock \bibinfo{journal}{\apjs} \bibinfo{volume}{250} (\bibinfo{year}{2020})
  \bibinfo{pages}{26}. \DOIprefix\doi{10.3847/1538-4365/abafc1}.
  \href{http://arxiv.org/abs/2008.03865}{{\tt arXiv:2008.03865}}.
%Type = Book
\bibitem[{Rasmussen and Williams(2006)}]{rasmussen2006gaussian}
\bibinfo{author}{C.~E. Rasmussen}, \bibinfo{author}{C.~K. Williams},
  \bibinfo{title}{Gaussian Processes for Machine Learning},
  \bibinfo{publisher}{MIT Press}, \bibinfo{year}{2006}.
%Type = Article
\bibitem[{{Seikel} et~al.(2012){Seikel}, {Clarkson}, and
  {Smith}}]{2012JCAP...06..036S}
\bibinfo{author}{M.~{Seikel}}, \bibinfo{author}{C.~{Clarkson}},
  \bibinfo{author}{M.~{Smith}},
\newblock \bibinfo{title}{{Reconstruction of dark energy and expansion dynamics
  using Gaussian processes}},
\newblock \bibinfo{journal}{\jcap} \bibinfo{volume}{2012}
  (\bibinfo{year}{2012}) \bibinfo{pages}{036}.
  \DOIprefix\doi{10.1088/1475-7516/2012/06/036}.
  \href{http://arxiv.org/abs/1204.2832}{{\tt arXiv:1204.2832}}.

\end{thebibliography}

\end{document}